\documentclass[twocolumn,twocolappendix]{aastex631}

\usepackage[toc,titletoc]{appendix}
\usepackage {appendix}
\usepackage[flushleft]{threeparttable}
\usepackage{xcolor,graphicx,xspace,color,longtable,enumitem}
\usepackage{amssymb,amsmath,shadow,bezier,curves,rotating}
\usepackage{graphicx,chngcntr}
\usepackage{multirow}
\graphicspath{ {./images/}}

\newcommand {\h}  {$h^{-1}\,$Mpc}

\newcommand {\m}  {M$_{\odot}$}

\newcommand {\M} {\mathcal{M}}
\newcommand {\N} {\mathcal{N}}

\begin{document}

\title[Environmental Quenching]{From Cluster Cores to the Low-Density Field: Strong Environmental Quenching of Galaxy Star Formation at Low Redshift}

\author[0000-0003-3595-7147]{Mohamed H. Abdullah}
\affiliation{Department of Physics, University of California Merced, 5200 North Lake Road, Merced, CA 95343, USA}
\affiliation{Department of Astronomy, National Research Institute of Astronomy and Geophysics, Cairo, 11421, Egypt}

\author[0000-0001-5071-2178]{A. E. Abdelaziz}
\affiliation{Department of Astronomy, National Research Institute of Astronomy and Geophysics, Cairo, 11421, Egypt}

\author[0000-0002-6572-7089]{Gillian Wilson}
\affiliation{Department of Physics, University of California Merced, 5200 North Lake Road, Merced, CA 95343, USA}

\author[0000-0002-5232-3368]{Ahmed M. Abdelbar}
\affiliation{Department of Astronomy and Meteorology, Faculty of Science, Al-Azhar University, Cairo, Egypt}

\author{M. M. Beheary}
\affiliation{Department of Astronomy and Meteorology, Faculty of Science, Al-Azhar University, Cairo, Egypt}

\author[0000-0002-2356-8315]{Y. H. M. Hendy}
\affiliation{Department of Astronomy, National Research Institute of Astronomy and Geophysics, Cairo, 11421, Egypt}

\begin{abstract}

We investigate how galaxy star formation activity depends on environment using a sample of 81,647 SDSS galaxies selected over $0.03\leq z\leq0.075$ and $9.7\leq\log_{10}(M_\star/h^{-2}M_\odot)\leq11.0$, including 18,426 members from 572 clusters in the \texttt{GalWCat19} catalog. We characterize environment in two complementary ways: (1) nearest-neighbor density for the full sample, and (2) clustercentric radius and host halo mass for \texttt{GalWCat19}. The sSFR distribution remains bimodal across all environments, with distinct quenched and star-forming components. As local density increases, the quenched component becomes more prominent, while the characteristic sSFR of the star-forming component decreases by approximately $0.29$--$0.35$ dex from the lowest- to highest-density classes. Within clusters, the quenched fraction decreases with increasing projected clustercentric radius, while the star-forming peak shifts by approximately $0.42$ dex toward lower sSFR from the outskirts to the inner cluster region. This extends the picture from previous studies, in which environmental trends are primarily associated with changes in the quenched fraction, by showing that galaxies remaining in the star-forming population also exhibit systematically suppressed sSFR in denser environments. The dependence on host halo mass is weaker and is most apparent among lower-stellar-mass galaxies in the inner cluster regions. By measuring the environmental quenching efficiency at fixed stellar mass, we find excess quenching in cluster environments beyond that expected from stellar-mass quenching alone. These results show that environment is associated not only with an increased probability of quenching, but also with suppressed star formation among galaxies that remain star forming, with local density and clustercentric radius showing the strongest associations.
\end{abstract}

\section{Introduction} \label{sec:intro}

A central goal of modern astrophysics is to understand how galaxies form and evolve across cosmic time and how actively star-forming systems become the diverse population of star-forming and quiescent galaxies observed today \citep{Naab17, Somerville15, Behroozi19}.
Nearby galaxies are the most evolved descendants of earlier galaxy populations, and their proximity enables precise measurements of their physical properties and environments. They therefore provide a detailed view of late-stage galaxy evolution and a key benchmark for theoretical models and simulations.

Galaxies in the local Universe exhibit a pronounced bimodality in color, star formation activity, and morphology \citep{Strateva01, Kauffmann03, Baldry04, Baldry06}. Star-forming, predominantly late-type galaxies populate the ``blue cloud,'' whereas quiescent, predominantly early-type galaxies occupy the ``red sequence'' and show strongly suppressed star formation \citep{Brinchmann04, Salim07, Bluck20}. The less populated ``green valley'' between these sequences is commonly associated with galaxies transitioning toward quiescence and therefore provides constraints on quenching pathways and timescales \citep{Schawinski14, Salim14, Tacchella15}. The star formation rate (SFR) and specific star formation rate (sSFR) are key diagnostics of galaxy evolution. Their distributions provide more information than a simple division between star-forming and quenched galaxies: they trace both the relative abundance of these populations and the suppression of star formation among galaxies that remain active. In clusters, quantitative measurements of the SFR and sSFR distributions as functions of clustercentric position and host halo mass provide empirical constraints on hydrodynamical simulations, semi-analytic models, and other theoretical descriptions of galaxy populations in dense environments.

Galaxy quenching is generally attributed to a combination of internal, mass-related processes and external, environmentally driven processes \citep{Peng10}. Internal quenching is associated with stellar-mass growth, supermassive black-hole activity, structural evolution, and regulation of the gas supply. Proposed mechanisms include active galactic nucleus feedback, morphological stabilization of gas disks, and exhaustion or heating of the available gas reservoir \citep{Croton06, Bower06, Fabian12, Martig09, Krumholz12}. The high quenched fractions of galaxies above approximately $\log(M_\star/M_\odot)\sim10.5$ across a broad range of environments indicate that internal processes become increasingly important toward high stellar masses \citep{Peng10, Thomas10}. Environmental quenching, by contrast, results from interactions between galaxies and their surroundings and is especially important for satellite galaxies in groups and clusters \citep{Dressler80, Bluck14}. Ram-pressure stripping may remove cold gas as galaxies move through the hot intracluster medium \citep{Gunn72, Abadi99, Boselli19}, while strangulation or starvation may terminate the supply of fresh gas by removing the extended gaseous halo \citep{Larson80, Balogh00, Kauffmann04, vandenBosch08}. Tidal interactions, repeated high-speed encounters, and mergers may also disturb galaxies, accelerate gas consumption, or trigger morphological transformation \citep{Moore96, Moore99, Hopkins08, Hopkins10}.

Large spectroscopic surveys have established that both stellar mass and environment influence galaxy star formation. Studies based on the 2dF Galaxy Redshift Survey and the Sloan Digital Sky Survey (SDSS; \citealp{York00, Alam15}) have linked galaxy properties to local density, group membership, and cluster environment \citep{Lewis02, Balogh04, DePropris04}. In particular, \citet{Peng10} showed that stellar mass and environment contribute separately to galaxy quenching. At fixed stellar mass, satellite galaxies in groups and clusters generally have higher quenched fractions than comparable field galaxies, with the environmental contribution being especially important for low- and intermediate-mass systems \citep{Peng10}. Within clusters, galaxies near the center typically have higher quenched fractions and lower SFRs and sSFRs than galaxies near or beyond the virial boundary \citep{Wetzel13, Wetzel15}. Host halo mass may introduce an additional dependence because more massive clusters generally contain hotter and denser intracluster media, although its relative importance compared with clustercentric radius remains uncertain \citep{Balogh11, Bluck16, Donnari21}. Environmental processing may also begin before galaxies enter the main cluster halo through pre-processing in groups, filaments, or infalling structures \citep{Fujita04, McGee09, Darvish17, Haines15}.

An important question is whether environment primarily changes the relative numbers of star-forming and quenched galaxies, or whether it also modifies the star formation activity of galaxies that remain star forming. \citet{Wetzel12,Wetzel13} found that environmental trends are dominated by changes in the quenched fraction, with comparatively little change in the star-forming population prior to quenching. This result underlies the delayed-then-rapid picture of satellite quenching. Measuring the full sSFR distribution as a function of environment therefore provides a means of testing whether environmental effects are confined primarily to the transition between star-forming and quenched populations, or are already apparent within the star-forming population itself.

To separate environmental quenching from stellar-mass quenching, galaxies must be compared at fixed stellar mass across different environments. A low-density reference population provides an estimate of the quenched fraction in the relative absence of strong cluster-related processes, while any excess quenching in clusters at the same stellar mass provides a practical measure of the additional environmental contribution \citep{Wetzel13, Cortese21}. In this work, we investigate how galaxy star formation activity depends on environment at low redshift using a homogeneous and approximately volume-complete sample of SDSS galaxies. We characterize environment in two complementary ways: projected nearest-neighbor surface densities for the full galaxy sample, and projected clustercentric radius and host halo mass for member galaxies of clusters in the \texttt{GalWCat19} catalog \citep{Abdullah20a} identified using the GalWeight technique \citep{Abdullah18}.

Our analysis addresses three main questions: 

how the mean SFR, mean sSFR, quenched fraction, and full sSFR distribution vary with local galaxy density, and in particular whether increasing density changes only the relative weights of the quenched and star-forming populations or also shifts the characteristic sSFR of galaxies that remain star forming; how the star formation properties of cluster members depend on projected clustercentric radius, host halo mass, and projected phase-space position; and whether cluster galaxies exhibit excess quenching beyond that expected from internal stellar-mass quenching alone. Extending the full sSFR-distribution decomposition to fixed stellar-mass intervals requires a substantially larger set of joint environment--stellar-mass subsamples. We therefore defer this analysis to a companion study. The resulting measurements provide quantitative observational constraints for models and simulations of the spatial and star formation distributions of galaxies in clusters.

The paper is organized as follows. Section~\ref{sec:data} describes the data, sample selection, galaxy properties, and environmental definitions. Section~\ref{sec:results} presents the dependence of star formation activity on local density and cluster environment, including the sSFR distributions, radial and halo-mass trends, projected phase-space analysis, and environmental quenching efficiency. In Section~\ref{sec:discussion}, we discuss the physical interpretation of these results, their connection to environmental quenching mechanisms, and their implications for models and simulations of cluster galaxy evolution. Throughout this paper, we adopt a flat $\Lambda$CDM cosmology consistent with the \citet{Planck15} results, assuming $\Omega_{\mathrm{M}} = 0.3089$, $\Omega_{\Lambda} = 0.6911$, and $h = 0.6774$. 
\section{Data, Galaxy Classification, and Environmental Definitions}
\label{sec:data}

In this section, we describe the observational sample, derived galaxy properties, sample-completeness criteria, and environmental definitions used to study the dependence of star formation activity on environment. Because our analysis compares star-forming and quenched galaxies across different environments, we adopt a restricted, approximately volume-complete sample to reduce population-dependent incompleteness in the flux-limited SDSS catalog. The parent galaxy sample is drawn from the Sloan Digital Sky Survey Data Release 13 (SDSS--DR13; \citealp{Albareti17}). This spectroscopic sample also provides the basis for the \texttt{GalWCat19} cluster catalog \citep{Abdullah20a}, which is used to define cluster membership, projected clustercentric radius, and host halo mass.

\subsection{Galaxy Physical Properties}
\label{sec:properties}

For each galaxy, we compile the physical quantities required for this analysis, including stellar mass ($M_\star$), star formation rate (SFR), and specific star formation rate (sSFR). These quantities are obtained by cross-matching the SDSS spectroscopic sample with the New York University Value-Added Galaxy Catalog (NYU-VAGC; SDSS--DR7; \citealp{Blanton05}) and the MPA/JHU value-added catalog\footnote{\url{http://www.mpa-garching.mpg.de/SDSS/DR7/}}.

The initial sample consists of spectroscopically confirmed SDSS galaxies identified by the automated SDSS classification pipeline. To ensure consistency in photometric and derived galaxy properties, we require each selected galaxy to have a counterpart in the NYU-VAGC. We adopt the standard SDSS main galaxy sample magnitude limits, $13.9 \leq r \leq 17.6$, in the Galactic-extinction-corrected Petrosian $r$ band \citep{Blanton03a}. We retain only galaxies with reliable spectroscopic redshifts, defined by \texttt{SpecObj.zWarning = 0}.

Star formation rates are based on the methodology of \citet{Brinchmann04}. Within the SDSS fiber aperture, SFRs are estimated from nebular emission-line diagnostics, particularly the $H_\alpha$ luminosity \citep{Kennicutt98}, with dust attenuation corrections based on the Balmer decrement \citep{Charlot00}. Total SFRs are then obtained by applying aperture corrections for star formation outside the fiber, using galaxy photometry following \citet{Salim07}. These measurements provide a homogeneous set of stellar masses and SFRs suitable for statistical comparisons of galaxy star formation activity across different environments.

\begin{figure*}\centering
\includegraphics[width=1\linewidth]{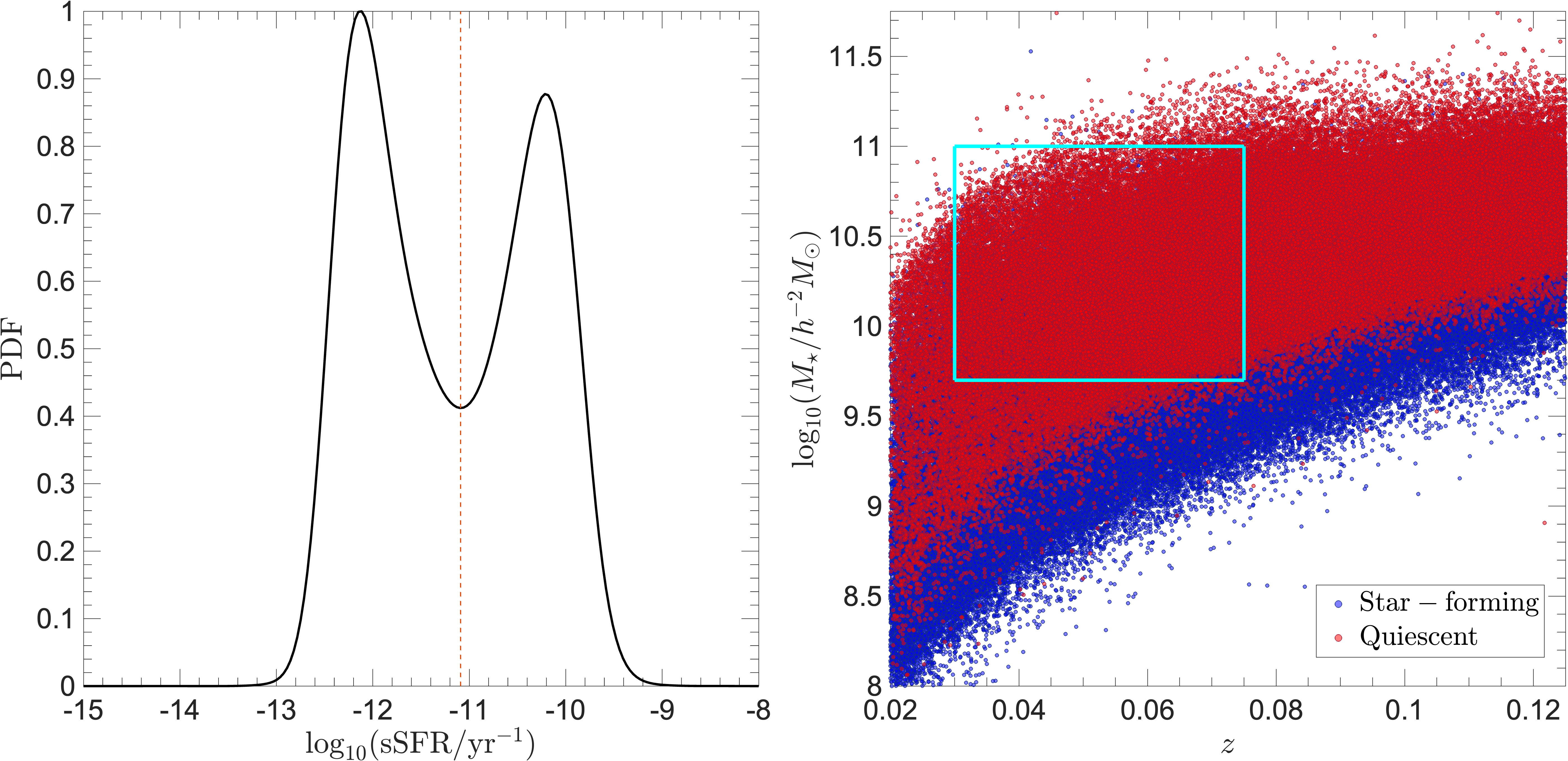}
\vspace{-0.5cm}
\caption{Galaxy classification and sample selection. Left: probability density function of $\log_{10}\mathrm{sSFR}$ for the galaxy sample. The vertical dashed line marks the adopted threshold, $\log_{10}\mathrm{sSFR} = -11.09$, used to separate quenched and star-forming galaxies.
Right: stellar mass--redshift distribution of the classified galaxies. The red and blue points indicate the quenched and star--forming galaxies. The cyan rectangle indicates the approximately volume-limited sample used in this work, defined by $0.03 \leq z \leq 0.075$ and $9.7 \leq \log_{10}(M_\star/h^{-2}M_{\odot}) \leq 11.0$.}
\label{fig:Comp_Class}
\end{figure*}
\subsection{Sample Completeness}
\label{sec:galclass}
A requirement for this analysis is to minimize the effects of sample incompleteness when comparing star-forming and quenched galaxies across different environments. In a flux-limited survey, stellar-mass completeness depends on galaxy population. At fixed redshift, star-forming galaxies can generally be detected to lower stellar masses than red quenched galaxies because of their lower mass-to-light ratios \citep[e.g.,][]{vandenBosch08,Mosleh2018}. This incomplete population can bias measurements of the quenched fraction and the sSFR distribution if low-mass star-forming galaxies are overrepresented relative to quenched systems. We therefore first classify galaxies into star-forming and quenched populations and then define a volume-limited stellar-mass and redshift range over which both populations can be compared consistently.

Galaxies are classified according to their specific star formation rate. The left panel of Figure~\ref{fig:Comp_Class} shows the probability density function of
$\log_{10}(\mathrm{sSFR}/\mathrm{yr}^{-1})$ for the parent galaxy sample. The distribution is bimodal, reflecting the separation between actively star-forming galaxies and quenched systems. We adopt the minimum between the two peaks as the division between the two populations. This minimum occurs at approximately $\log_{10}(\mathrm{sSFR}/\mathrm{yr}^{-1})=-11.09$. Galaxies with $\log_{10}(\mathrm{sSFR}/\mathrm{yr}^{-1}) > -11.09$ are classified as star-forming, while galaxies with $\log_{10}(\mathrm{sSFR}/\mathrm{yr}^{-1}) \leq -11.09$ are classified as quenched.

Figure~\ref{fig:Comp_Class} (right) shows the stellar mass-redshift distribution of the classified galaxies. The cyan rectangle marks the approximately volume-limited sample adopted in this work, defined by $0.03 \leq z \leq 0.075$ and $9.7 \leq \log_{10}(M_\star/h^{-2}M_{\odot}) \leq 11.0$. This selection reduces redshift-dependent incompleteness and avoids population-dependent biases that can affect comparisons between quenched and star-forming galaxies at low stellar masses. The relatively low upper-redshift limit allows us to retain galaxies down to lower stellar masses while minimizing evolutionary effects across the sample. The resulting sample contains $81,647$ galaxies.

Although the adopted stellar-mass and redshift limits produce an approximately volume-complete sample, residual differences remain in the comoving volume over which individual galaxies satisfy the SDSS apparent-magnitude limits. We therefore assign each galaxy a weight proportional to $1/V_{\max}$ \citep[e.g.,][]{Weigel16} in all binned measurements and distribution fits. Here, $V_{\max}$ is the maximum accessible comoving volume over which the galaxy would remain within the survey selection limits. 
The restricted stellar-mass and redshift selection, together with the $1/V_{\max}$ weighting, therefore reduces residual selection-related biases in the measured SFR, sSFR, and quenched fraction.
\begin{table*}
\centering
\caption{Summary of the environmental classifications and corresponding galaxy sample sizes used in this work.}
\label{tab:env_classes} \tabletypesize{\scriptsize} \setlength{\tabcolsep}{4pt}
\begin{tabular}{lllr}
\hline
\hline
Environment category & Class & Definition & $N_{\rm gal}$ \\

\hline
\multicolumn{4}{l}{\textit{Statistical-density environments, $N=5$}} \\
Lowest-density class& $C_{5,1}$& $\Sigma_5<P_{20}(\Sigma_5)$& 16329 \\
Intermediate-low-density class& $C_{5,2}$& $P_{20}(\Sigma_5)\leq\Sigma_5<P_{50}(\Sigma_5)$& 24493 \\
Intermediate-high-density class& $C_{5,3}$& $P_{50}(\Sigma_5)\leq\Sigma_5<P_{80}(\Sigma_5)$& 24496 \\
Highest-density class& $C_{5,4}$& $\Sigma_5\geq P_{80}(\Sigma_5)$& 16329 \\

\hline
\multicolumn{4}{l}{\textit{Statistical-density environments, $N=10$}} \\
Lowest-density class& $C_{10,1}$& $\Sigma_{10}<P_{20}(\Sigma_{10})$& 16329 \\
Intermediate-low-density class& $C_{10,2}$& $P_{20}(\Sigma_{10})\leq\Sigma_{10}<P_{50}(\Sigma_{10})$& 24493 \\
Intermediate-high-density class& $C_{10,3}$& $P_{50}(\Sigma_{10})\leq\Sigma_{10}<P_{80}(\Sigma_{10})$& 24496 \\
Highest-density class& $C_{10,4}$& $\Sigma_{10}\geq P_{80}(\Sigma_{10})$& 16329 \\

\hline
\multicolumn{4}{l}{\textit{Statistical-density environments, $N=15$}} \\
Lowest-density class& $C_{15,1}$& $\Sigma_{15}<P_{20}(\Sigma_{15})$& 16329 \\
Intermediate-low-density class& $C_{15,2}$& $P_{20}(\Sigma_{15})\leq\Sigma_{15}<P_{50}(\Sigma_{15})$& 24493 \\
Intermediate-high-density class& $C_{15,3}$& $P_{50}(\Sigma_{15})\leq\Sigma_{15}<P_{80}(\Sigma_{15})$& 24496 \\
Highest-density class& $C_{15,4}$& $\Sigma_{15}\geq P_{80}(\Sigma_{15})$& 16329 \\

\hline
\multicolumn{4}{l}{\textit{Cluster-scale radial environments}} \\
All cluster members& AllMem& All \texttt{GalWCat19} cluster members& 18426 \\
Inner cluster region& $R_1$& $R_{\rm p}/R_{200}<0.5$& 3776 \\
Outer virial region& $R_2$& $0.5\leq R_{\rm p}/R_{200}<1$& 2927 \\
Cluster outskirts and infall region& $R_3$& $1\leq R_{\rm p}/R_{200}\leq6$& 11720 \\

\hline
\multicolumn{4}{l}{\textit{Cluster halo-mass environments}} \\
Low-mass halos& $M_1$& $\log_{10}(M_{200}/h^{-1}M_{\odot})<14.1$& 9808 \\
Intermediate-mass halos& $M_2$& $14.1\leq\log_{10}(M_{200}/h^{-1}M_{\odot})<14.5$& 6182 \\
High-mass halos& $M_3$& $\log_{10}(M_{200}/h^{-1}M_{\odot})\geq14.5$& 2436 \\

\hline
\multicolumn{4}{l}{\textit{Joint radial and halo-mass environments}} \\
Inner region, low-mass halos& $R_1M_1$& $R_1\cap M_1$& 2114 \\
Inner region, intermediate-mass halos& $R_1M_2$& $R_1\cap M_2$& 1219 \\
Inner region, high-mass halos& $R_1M_3$& $R_1\cap M_3$& 443 \\
Intermediate region, low-mass halos& $R_2M_1$& $R_2\cap M_1$& 1562 \\
Intermediate region, intermediate-mass halos& $R_2M_2$& $R_2\cap M_2$& 976 \\
Intermediate region, high-mass halos& $R_2M_3$& $R_2\cap M_3$& 389 \\
Outer region, low-mass halos& $R_3M_1$& $R_3\cap M_1$& 6129 \\
Outer region, intermediate-mass halos& $R_3M_2$& $R_3\cap M_2$& 3987 \\
Outer region, high-mass halos& $R_3M_3$& $R_3\cap M_3$& 1604 \\
\hline
\end{tabular}
\tablecomments{The projected nearest-neighbor density is defined as $\Sigma_N=N/(\pi d_N^2)$, where $d_N$ is the projected distance to the $N$th nearest neighbor within the adopted line-of-sight velocity interval. For each value of $N$, $P_{20}(\Sigma_N)$, $P_{50}(\Sigma_N)$, and $P_{80}(\Sigma_N)$ denote the 20th, 50th, and 80th percentiles of the corresponding $\Sigma_N$ distribution. The classes
$C_{N,1}$--$C_{N,4}$ are ordered from the lowest- to highest-density environments. The joint class $R_iM_j$ contains galaxies that satisfy both the radial definition $R_i$ and the halo-mass definition $M_j$. The AllMem and $M_i$ samples include all applicable cluster members, whereas the radial and joint samples are restricted to $R_{\rm p}/R_{200}\leq6$.}
\end{table*}

\subsection{Large-Scale Environment Classification}
\label{sec:env}

To investigate how galaxy star formation activity depends on large-scale environment, we characterize the local density surrounding each galaxy using a projected nearest-neighbor density estimator. This approach provides a continuous measure of the local galaxy density and can be applied uniformly across the full sample without requiring the prior identification of specific structures such as clusters or groups.

For each galaxy, we calculate the projected comoving distance to its $N$-th nearest neighbor, $d_N$, considering neighboring galaxies within a line-of-sight velocity interval of $\pm2000~\mathrm{km\,s^{-1}}$. The tracer population includes all galaxies in the adopted sample, including both quenched and star-forming systems. The corresponding projected surface density is defined as
\begin{equation}
\Sigma_N = \frac{N}{\pi d_N^2}.
\end{equation}
We compute this estimator for $N=5$, 10, and 15 in order to probe surrounding galaxy density over slightly different smoothing scales. For each value of $N$, galaxies are divided into four statistical density classes using the 20th, 50th, and 80th percentiles of the corresponding $\Sigma_N$ distribution. We denote these classes as $C_{N,1}$, $C_{N,2}$, $C_{N,3}$, and $C_{N,4}$, where the first subscript specifies the nearest-neighbor order and the second subscript specifies the density class. For a given $N$, $C_{N,1}$ represents the lowest-density environment, $C_{N,4}$ represents the highest-density environment, and $C_{N,2}$ and $C_{N,3}$ represent the intermediate-low- and intermediate-high- density environments, respectively.

These statistical density classes differ conceptually from the cluster-based environments introduced below. The nearest-neighbor estimator ranks galaxies according to their local projected density without explicitly identifying gravitationally bound halos. Consequently, the higher-density classes may contain galaxies associated with clusters, groups, filaments, and  cluster outskirts and infall region. In contrast, the cluster-based environments are defined using physically identified systems and galaxy membership assignments from \texttt{GalWCat19}. We verified that the fiducial lowest-density class, $C_{10,1}$, contains no \texttt{GalWCat19} members. We therefore adopt it as the low-density non-cluster reference population for the environmental quenching analysis.

\begin{figure*}
\centering 
\includegraphics[width=1\linewidth]{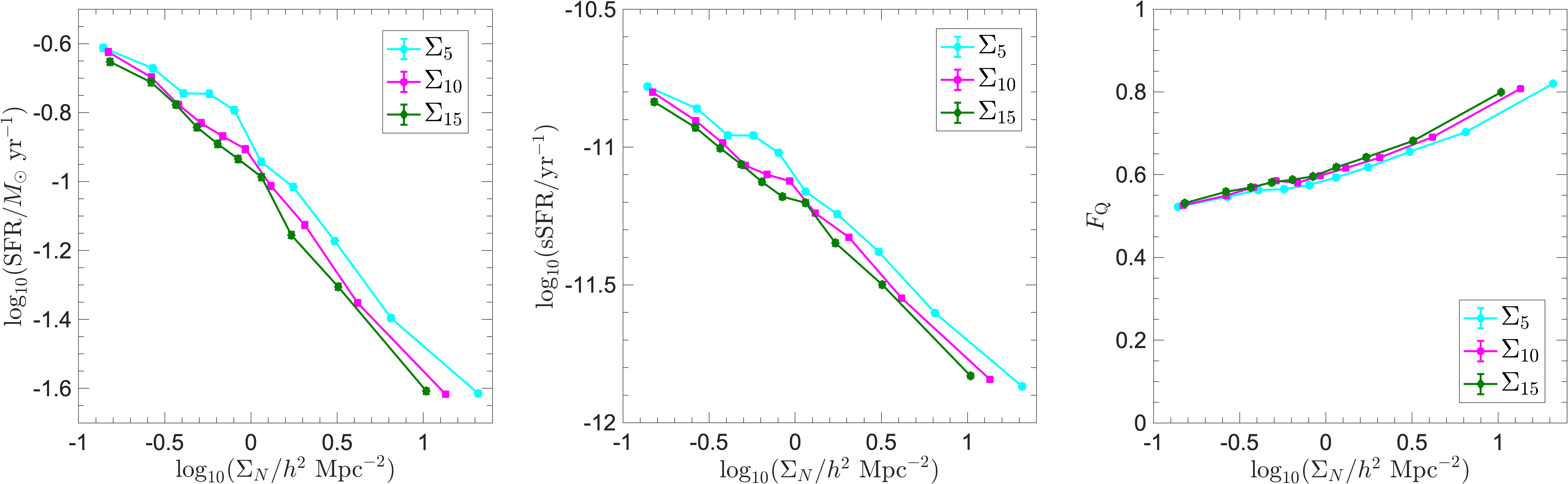} \vspace{-0.5cm} 
\caption{
Environmental dependence of star formation activity as a function of the projected nearest-neighbor surface density, $\Sigma_N$. From left to right, the panels show the mean $\log_{10}(\mathrm{SFR}/M_\odot~\mathrm{yr}^{-1})$, the mean $\log_{10}(\mathrm{sSFR}/\mathrm{yr}^{-1})$, and the quenched fraction, $F_{\rm Q}$, as functions of $\log_{10}(\Sigma_N/h^2~\mathrm{Mpc}^{-2})$. Results are shown for $N=5$, 10, and 15.
} 
\label{fig:SigmaEnv} 
\end{figure*}

\begin{table*}
\centering
\caption{Weighted quenched fractions and best-fitting weighted two-component Gaussian-mixture parameters for the $\log_{10}(\mathrm{sSFR}/\mathrm{yr}^{-1})$ distributions across the statistical-density and cluster environments.}
\label{tab:gmm_env}
\tabletypesize{\scriptsize}
\setlength{\tabcolsep}{2.5pt}
\begin{tabular}{lcccccc}
\hline
\hline
Environment & $F_{\rm Q,w}$ & $\mu_{\rm Q}$ & $\sigma_{\rm Q}$ & $w_{\rm Q}$ & $\mu_{\rm SF}$ & $\sigma_{\rm SF}$ \\
\hline
\multicolumn{7}{l}{\textit{Statistical-density environments, $N=5$}} \\
$C_{5,1}$ & $0.441_{-0.004}^{+0.004}$ & $-11.852_{-0.008}^{+0.014}$ & $0.418_{-0.006}^{+0.010}$ & $0.448_{-0.005}^{+0.008}$ & $-10.297_{-0.006}^{+0.011}$ & $0.375_{-0.008}^{+0.005}$ \\
$C_{5,2}$ & $0.478_{-0.003}^{+0.003}$ & $-11.907_{-0.006}^{+0.012}$ & $0.403_{-0.004}^{+0.009}$ & $0.470_{-0.004}^{+0.007}$ & $-10.338_{-0.006}^{+0.011}$ & $0.393_{-0.008}^{+0.004}$ \\
$C_{5,3}$ & $0.539_{-0.003}^{+0.003}$ & $-11.971_{-0.007}^{+0.013}$ & $0.369_{-0.004}^{+0.009}$ & $0.499_{-0.004}^{+0.009}$ & $-10.428_{-0.007}^{+0.016}$ & $0.444_{-0.011}^{+0.005}$ \\
$C_{5,4}$ & $0.717_{-0.004}^{+0.004}$ & $-12.029_{-0.012}^{+0.008}$ & $0.332_{-0.009}^{+0.005}$ & $0.633_{-0.015}^{+0.008}$ & $-10.647_{-0.038}^{+0.020}$ & $0.535_{-0.012}^{+0.022}$ \\
\hline
\multicolumn{7}{l}{\textit{Statistical-density environments, $N=10$}} \\
$C_{10,1}$ & $0.448_{-0.004}^{+0.004}$ & $-11.853_{-0.009}^{+0.017}$ & $0.424_{-0.006}^{+0.011}$ & $0.456_{-0.005}^{+0.009}$ & $-10.297_{-0.006}^{+0.012}$ & $0.375_{-0.009}^{+0.005}$ \\
$C_{10,2}$ & $0.493_{-0.003}^{+0.004}$ & $-11.910_{-0.010}^{+0.007}$ & $0.402_{-0.007}^{+0.005}$ & $0.486_{-0.006}^{+0.005}$ & $-10.335_{-0.009}^{+0.006}$ & $0.392_{-0.005}^{+0.007}$ \\
$C_{10,3}$ & $0.532_{-0.003}^{+0.003}$ & $-11.962_{-0.010}^{+0.009}$ & $0.378_{-0.007}^{+0.006}$ & $0.498_{-0.007}^{+0.007}$ & $-10.416_{-0.012}^{+0.010}$ & $0.433_{-0.007}^{+0.008}$ \\
$C_{10,4}$ & $0.700_{-0.003}^{+0.004}$ & $-12.022_{-0.013}^{+0.007}$ & $0.332_{-0.009}^{+0.004}$ & $0.620_{-0.014}^{+0.008}$ & $-10.624_{-0.032}^{+0.018}$ & $0.528_{-0.010}^{+0.020}$ \\
\hline
\multicolumn{7}{l}{\textit{Statistical-density environments, $N=15$}} \\
$C_{15,1}$ & $0.452_{-0.004}^{+0.004}$ & $-11.857_{-0.008}^{+0.014}$ & $0.418_{-0.005}^{+0.010}$ & $0.458_{-0.004}^{+0.009}$ & $-10.305_{-0.006}^{+0.011}$ & $0.378_{-0.009}^{+0.004}$ \\
$C_{15,2}$ & $0.498_{-0.003}^{+0.003}$ & $-11.913_{-0.010}^{+0.008}$ & $0.402_{-0.007}^{+0.005}$ & $0.489_{-0.006}^{+0.005}$ & $-10.340_{-0.009}^{+0.008}$ & $0.396_{-0.005}^{+0.007}$ \\
$C_{15,3}$ & $0.533_{-0.003}^{+0.003}$ & $-11.964_{-0.010}^{+0.007}$ & $0.378_{-0.007}^{+0.005}$ & $0.502_{-0.007}^{+0.005}$ & $-10.411_{-0.012}^{+0.009}$ & $0.429_{-0.006}^{+0.008}$ \\
$C_{15,4}$ & $0.686_{-0.004}^{+0.004}$ & $-12.014_{-0.012}^{+0.007}$ & $0.335_{-0.009}^{+0.004}$ & $0.612_{-0.015}^{+0.007}$ & $-10.595_{-0.033}^{+0.015}$ & $0.519_{-0.009}^{+0.019}$ \\
\hline
\multicolumn{7}{l}{\textit{Cluster-scale radial environments}} \\
AllMem & $0.655_{-0.006}^{+0.005}$ & $-12.020_{-0.012}^{+0.007}$ & $0.332_{-0.008}^{+0.004}$ & $0.580_{-0.013}^{+0.006}$ & $-10.572_{-0.025}^{+0.014}$ & $0.511_{-0.008}^{+0.016}$ \\
$R_1$ & $0.811_{-0.009}^{+0.008}$ & $-12.085_{-0.012}^{+0.017}$ & $0.297_{-0.009}^{+0.012}$ & $0.663_{-0.022}^{+0.030}$ & $-10.941_{-0.052}^{+0.077}$ & $0.663_{-0.029}^{+0.016}$ \\
$R_2$ & $0.717_{-0.010}^{+0.011}$ & $-11.993_{-0.022}^{+0.016}$ & $0.327_{-0.013}^{+0.010}$ & $0.638_{-0.027}^{+0.019}$ & $-10.646_{-0.062}^{+0.046}$ & $0.538_{-0.033}^{+0.040}$ \\
$R_3$ & $0.586_{-0.006}^{+0.006}$ & $-12.008_{-0.014}^{+0.010}$ & $0.341_{-0.010}^{+0.007}$ & $0.518_{-0.012}^{+0.007}$ & $-10.522_{-0.021}^{+0.016}$ & $0.485_{-0.009}^{+0.014}$ \\
\hline
\multicolumn{7}{l}{\textit{Cluster halo-mass environments}} \\
$M_1$ & $0.636_{-0.006}^{+0.006}$ & $-12.019_{-0.015}^{+0.010}$ & $0.336_{-0.010}^{+0.006}$ & $0.564_{-0.014}^{+0.010}$ & $-10.568_{-0.031}^{+0.019}$ & $0.504_{-0.011}^{+0.017}$ \\
$M_2$ & $0.673_{-0.009}^{+0.009}$ & $-12.037_{-0.013}^{+0.017}$ & $0.316_{-0.009}^{+0.011}$ & $0.581_{-0.014}^{+0.018}$ & $-10.610_{-0.030}^{+0.039}$ & $0.540_{-0.025}^{+0.018}$ \\
$M_3$ & $0.684_{-0.016}^{+0.015}$ & $-11.989_{-0.030}^{+0.023}$ & $0.347_{-0.016}^{+0.013}$ & $0.632_{-0.023}^{+0.017}$ & $-10.524_{-0.056}^{+0.047}$ & $0.489_{-0.025}^{+0.032}$ \\
\hline
\multicolumn{7}{l}{\textit{Joint radial and halo-mass environments}} \\
$R_1M_1$ & $0.782_{-0.011}^{+0.011}$ & $-12.064_{-0.028}^{+0.024}$ & $0.320_{-0.020}^{+0.015}$ & $0.679_{-0.046}^{+0.032}$ & $-10.787_{-0.116}^{+0.095}$ & $0.587_{-0.046}^{+0.051}$ \\
$R_1M_2$ & $0.838_{-0.015}^{+0.014}$ & $-12.104_{-0.021}^{+0.024}$ & $0.277_{-0.014}^{+0.014}$ & $0.659_{-0.035}^{+0.040}$ & $-11.063_{-0.067}^{+0.089}$ & $0.703_{-0.032}^{+0.019}$ \\
$R_1M_3$ & $0.871_{-0.016}^{+0.019}$ & $-12.055_{-0.040}^{+0.033}$ & $0.294_{-0.022}^{+0.017}$ & $0.731_{-0.048}^{+0.058}$ & $-11.035_{-0.117}^{+0.163}$ & $0.717_{-0.044}^{+0.015}$ \\
$R_2M_1$ & $0.691_{-0.013}^{+0.013}$ & $-11.993_{-0.027}^{+0.021}$ & $0.328_{-0.017}^{+0.013}$ & $0.615_{-0.029}^{+0.023}$ & $-10.629_{-0.071}^{+0.055}$ & $0.510_{-0.031}^{+0.034}$ \\
$R_2M_2$ & $0.730_{-0.018}^{+0.016}$ & $-12.006_{-0.029}^{+0.034}$ & $0.321_{-0.021}^{+0.023}$ & $0.634_{-0.046}^{+0.048}$ & $-10.682_{-0.098}^{+0.108}$ & $0.580_{-0.096}^{+0.089}$ \\
$R_2M_3$ & $0.795_{-0.027}^{+0.023}$ & $-11.958_{-0.054}^{+0.049}$ & $0.334_{-0.029}^{+0.021}$ & $0.745_{-0.060}^{+0.035}$ & $-10.619_{-0.207}^{+0.154}$ & $0.550_{-0.080}^{+0.090}$ \\
$R_3M_1$ & $0.570_{-0.008}^{+0.007}$ & $-12.014_{-0.021}^{+0.016}$ & $0.339_{-0.014}^{+0.011}$ & $0.499_{-0.017}^{+0.014}$ & $-10.529_{-0.029}^{+0.027}$ & $0.490_{-0.016}^{+0.019}$ \\
$R_3M_2$ & $0.605_{-0.010}^{+0.010}$ & $-12.016_{-0.022}^{+0.021}$ & $0.328_{-0.016}^{+0.013}$ & $0.528_{-0.020}^{+0.017}$ & $-10.537_{-0.037}^{+0.033}$ & $0.497_{-0.021}^{+0.023}$ \\
$R_3M_3$ & $0.603_{-0.017}^{+0.013}$ & $-11.975_{-0.033}^{+0.033}$ & $0.368_{-0.024}^{+0.027}$ & $0.560_{-0.027}^{+0.021}$ & $-10.475_{-0.048}^{+0.054}$ & $0.451_{-0.025}^{+0.025}$ \\
\hline
\end{tabular}
\tablecomments{
$F_{\rm Q,w}$ is the $1/V_{\max}$-weighted quenched fraction, calculated using the adopted threshold $\log_{10}(\mathrm{sSFR}/\mathrm{yr}^{-1})=-11.09$. The Gaussian components are ordered by their fitted mean sSFR, such that the lower-sSFR and higher-sSFR components are labeled Q and SF, respectively. The quantity $w_{\rm Q}$ is the fitted mixture weight of the Q component and need not be identical to $F_{\rm Q,w}$. The parameters $\mu_{\rm Q}$ and $\sigma_{\rm Q}$ are the mean and dispersion of the lower-sSFR Gaussian component, while $\mu_{\rm SF}$ and $\sigma_{\rm SF}$ are the corresponding quantities for the higher-sSFR Gaussian component. The lower and upper uncertainties are obtained from the 16th and 84th percentiles of 1000 bootstrap realizations. Environment definitions and sample sizes are given in Table~\ref{tab:env_classes}.
}
\end{table*}

\subsection{Cluster-Scale Environmental Definitions}
\label{sec:cluster_env}

In addition to the statistical large-scale environments defined above, we characterize the environments of galaxies associated with \texttt{GalWCat19} clusters. These cluster-scale definitions allow us to examine how star formation activity varies with projected position within and around clusters, and with the mass of the host halo. 

We restrict the cluster sample to the same redshift interval adopted for the parent galaxy sample, $0.03 \leq z \leq 0.075$. The resulting sample contains $572$ clusters spanning the approximate dynamical-mass range $13.6 \leq\log_{10}(M_{200}/h^{-1}M_{\odot})\leq 15.1$, and includes $18,426$ cluster members. For each cluster, $R_{200}$ is defined as the radius within which the mean enclosed density is $200$ times the critical density of the Universe at the cluster redshift, $\rho_{\rm c}(z)$. The corresponding enclosed mass, $M_{200}$, is therefore given by
\begin{equation}
M_{200}=\frac{4\pi}{3}\,200\,\rho_{\rm c}(z)\,R_{200}^{3}.
\end{equation}
The quantity $\sigma_{200}$ denotes the rest-frame line-of-sight velocity dispersion of the galaxies associated with the cluster and is used to normalize galaxy peculiar velocities in the projected phase-space analysis. The rest-frame line-of-sight peculiar velocity is calculated as
\begin{equation}
V_{\rm pec}=c\,\frac{z-z_{\rm cl}}{1+z_{\rm cl}},
\end{equation}
where $z$ and $z_{\rm cl}$ are the galaxy and cluster redshifts,
respectively.

For each cluster member, we calculate the projected clustercentric distance, $R_{\rm p}$, relative to the cluster center provided by \texttt{GalWCat19}. We normalize this distance by $R_{200}$, producing a dimensionless measure of galaxy position that allows clusters of different physical sizes to be combined in a stacked analysis. The cluster members are divided into three radial classes, $R_1$, $R_2$, and $R_3$, corresponding to the inner cluster
region, the outer virial region, and the cluster outskirts and infall region, respectively. We also examine the dependence of galaxy star formation on host halo mass by dividing the clusters into three classes, $M_1$, $M_2$, and $M_3$, representing low-, intermediate-, and high-mass halos, respectively.

To separate the effects of clustercentric radius and host halo mass, we further construct nine joint environmental classes, $R_iM_j$, by combining each radial class with each halo-mass class. For example, $R_1M_1$ contains galaxies in the inner regions of low-mass halos, whereas $R_3M_3$ contains galaxies in the outskirts of high-mass halos. These joint classes allow us to examine radial variations in the sSFR distribution at fixed host halo mass and halo-mass variations at fixed clustercentric radius. The definitions and galaxy sample sizes of all statistical-density, radial, halo-mass, and joint environmental classes
are summarized in Table~\ref{tab:env_classes}.
\subsection{sSFR Peak Measurements}
\label{sec:ssfr_peaks}

To quantify the full star formation distribution in each environment, we model the weighted $\log_{10}(\mathrm{sSFR})$ distribution using a two-component Gaussian mixture model. The model is written as
\begin{equation}
P(x) =
w_{\rm Q}\,G(x|\mu_{\rm Q},\sigma_{\rm Q})
+
w_{\rm SF}\,G(x|\mu_{\rm SF},\sigma_{\rm SF}),
\end{equation}
where $x=\log_{10}(\mathrm{sSFR}/\mathrm{yr}^{-1})$. The lower-sSFR component is associated with the quenched population, while the higher-sSFR component is associated with the star-forming population.

For each environment class, the fit returns the peak location, width, and relative weight of each component, denoted by $(\mu_{\rm Q},\sigma_{\rm Q},w_{\rm Q})$ and $(\mu_{\rm SF},\sigma_{\rm SF},w_{\rm SF})$. The peak locations $\mu_{\rm Q}$ and $\mu_{\rm SF}$ represent the characteristic $\log_{10}(\mathrm{sSFR})$ values of the quenched and star-forming components, while $\sigma_{\rm Q}$ and $\sigma_{\rm SF}$ describe their widths. The mixture weights $w_{\rm Q}$ and $w_{\rm SF}$ represent the weighted contribution of each component to the total distribution, with $w_{\rm Q}+w_{\rm SF}=1$. All Gaussian-mixture fits are performed using the adopted $1/V_{\max}$ galaxy weights.
\section{Results}
\label{sec:results}
In this section, we study how star formation activity varies depending on galaxy environment. We consider statistical large-scale environments based on the projected nearest-neighbor density $\Sigma_N$, together with the \texttt{GalWCat19} cluster-scale environments based on projected clustercentric radius and host halo mass.

\begin{figure*}
\centering
\includegraphics[width=1\linewidth]{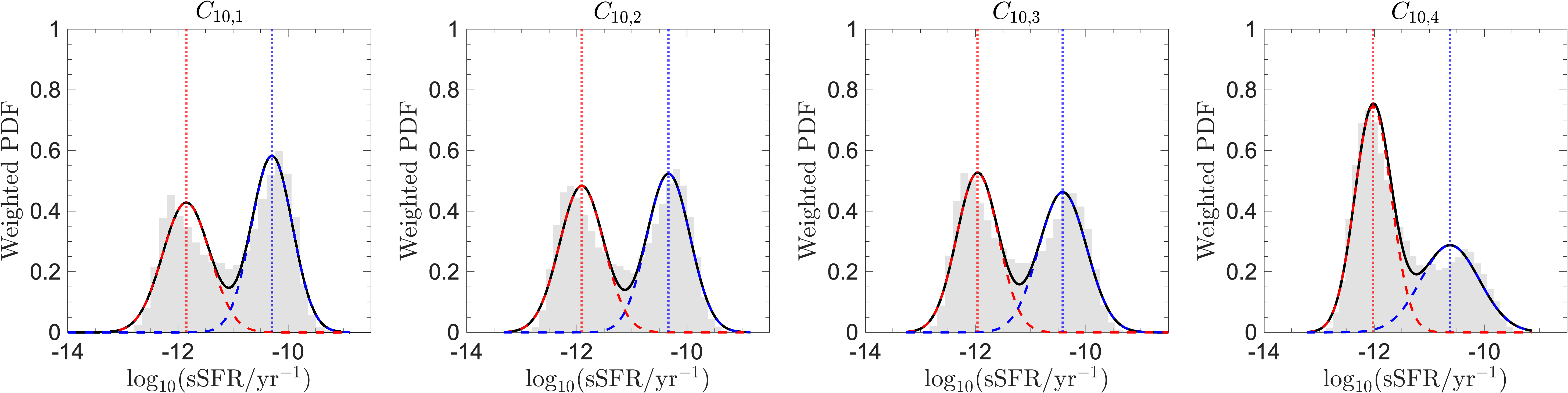} 
    \vspace{-0.5cm}
\caption{
Weighted distributions of $\log_{10}(\mathrm{sSFR}/\mathrm{yr}^{-1})$ for the four statistical density environments defined using $\Sigma_{10}$. Gray histograms show the weighted observed distributions. Black solid curves show the total two-component Gaussian mixture model. Red and blue dashed curves show the quenched and star-forming Gaussian components, respectively. Vertical dotted lines mark the peak positions $\mu_{\rm Q}$ and $\mu_{\rm SF}$. The panels are ordered from the lowest-density class, $C_{10,1}$, to the highest-density class, $C_{10,4}$.
}
\label{fig:DistLS}
\end{figure*}
\subsection{Star Formation Activity as a Function of Statistical Density}
\label{sec:results_ssfr_sigma}

We first examine the continuous dependence of star formation activity on the statistical large-scale environment. Figure~\ref{fig:SigmaEnv} shows the weighted mean $\log_{10}(\mathrm{SFR}/M_\odot~\mathrm{yr}^{-1})$, the weighted mean $\log_{10}(\mathrm{sSFR}/\mathrm{yr}^{-1})$, and the weighted quenched fraction, $F_{\rm Q}$, as functions of the projected nearest-neighbor surface density, $\Sigma_N$, for $N=5$, 10, and 15. These three choices probe slightly different environmental scales and allow us to test whether the measured trends depend on the adopted nearest-neighbor order. The left panel shows that the weighted mean $\log_{10}(\mathrm{SFR})$ decreases with increasing $\Sigma_N$. This indicates that galaxies in denser regions have lower absolute star formation rates. However, because SFR is strongly correlated with stellar mass, the sSFR provides a more direct measure of star formation activity relative to the already assembled stellar mass. The middle panel shows that the weighted mean $\log_{10}(\mathrm{sSFR}/\mathrm{yr}^{-1})$ also decreases monotonically with increasing density for all three values of $N$. Thus, galaxies in denser environments are not only forming fewer stars, but are also growing less efficiently relative to their stellar mass.

The right panel shows the corresponding behavior of $F_{\rm Q}$. The fraction of quenched galaxies increases systematically with $\Sigma_N$, rising from $F_{\rm Q}\simeq 0.5$ in the lowest-density bins to $F_{\rm Q}\gtrsim 0.8$ in the highest-density bins. This trend is consistent with the decline in the weighted mean $\log_{10}(\mathrm{sSFR})$ and shows that dense environments contain a larger fraction of galaxies with strongly suppressed star formation. The results are highly consistent for $N=5$, 10, and 15. The small offsets among the three curves reflect the fact that different nearest-neighbor orders probe slightly different smoothing scales, but the overall environmental dependence is unchanged. This agreement indicates that the observed suppression of star formation is not driven by a particular choice of density estimator.
\subsection{The sSFR Distribution Across Statistical Density Classes}
\label{sec:DistLS}
We next examine how the full $\log_{10}(\mathrm{sSFR})$ distribution changes across the statistical density classes. For each nearest-neighbor estimator, galaxies are divided into four classes, $C_{N,1}$--$C_{N,4}$, ordered from the lowest- to the highest-density environments (Table~\ref{tab:env_classes}). We fit the weighted $\log_{10}(\mathrm{sSFR})$ distribution in each class with a two-component Gaussian mixture model (Section \ref{sec:ssfr_peaks}). The resulting peak locations, widths, and mixture weights are listed in Table~\ref{tab:gmm_env}.
\begin{figure*}
\centering
\includegraphics[width=1\linewidth]{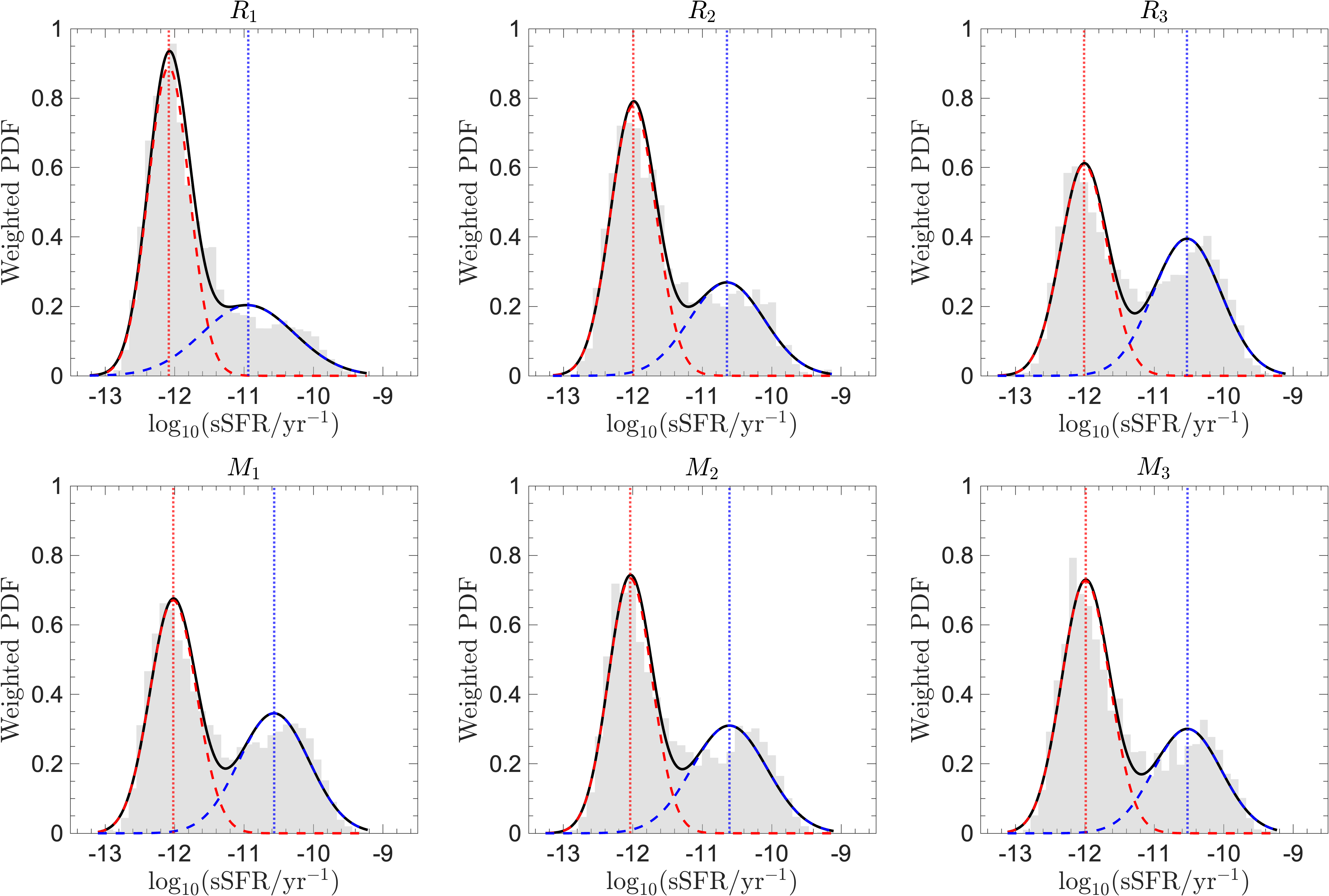} 
    \vspace{-0.5cm}
\caption{
Weighted $\log_{10}(\mathrm{sSFR}/\mathrm{yr}^{-1})$ distributions for cluster-scale environment samples. The top row shows the radial samples $R_1$--$R_3$, ordered from the inner cluster region to the cluster outskirts and infall region. The bottom row shows the host halo-mass samples $M_1$--$M_3$. Gray histograms show the weighted observed distributions. Black solid curves show the total two-component Gaussian mixture model. Red and blue dashed curves show the quenched and star-forming Gaussian components, respectively. Vertical dotted lines mark the corresponding peak positions, $\mu_{\rm Q}$ and $\mu_{\rm SF}$.
}
\label{fig:DistSS}
\end{figure*}
\begin{figure*}
\centering
\includegraphics[width=1\linewidth]{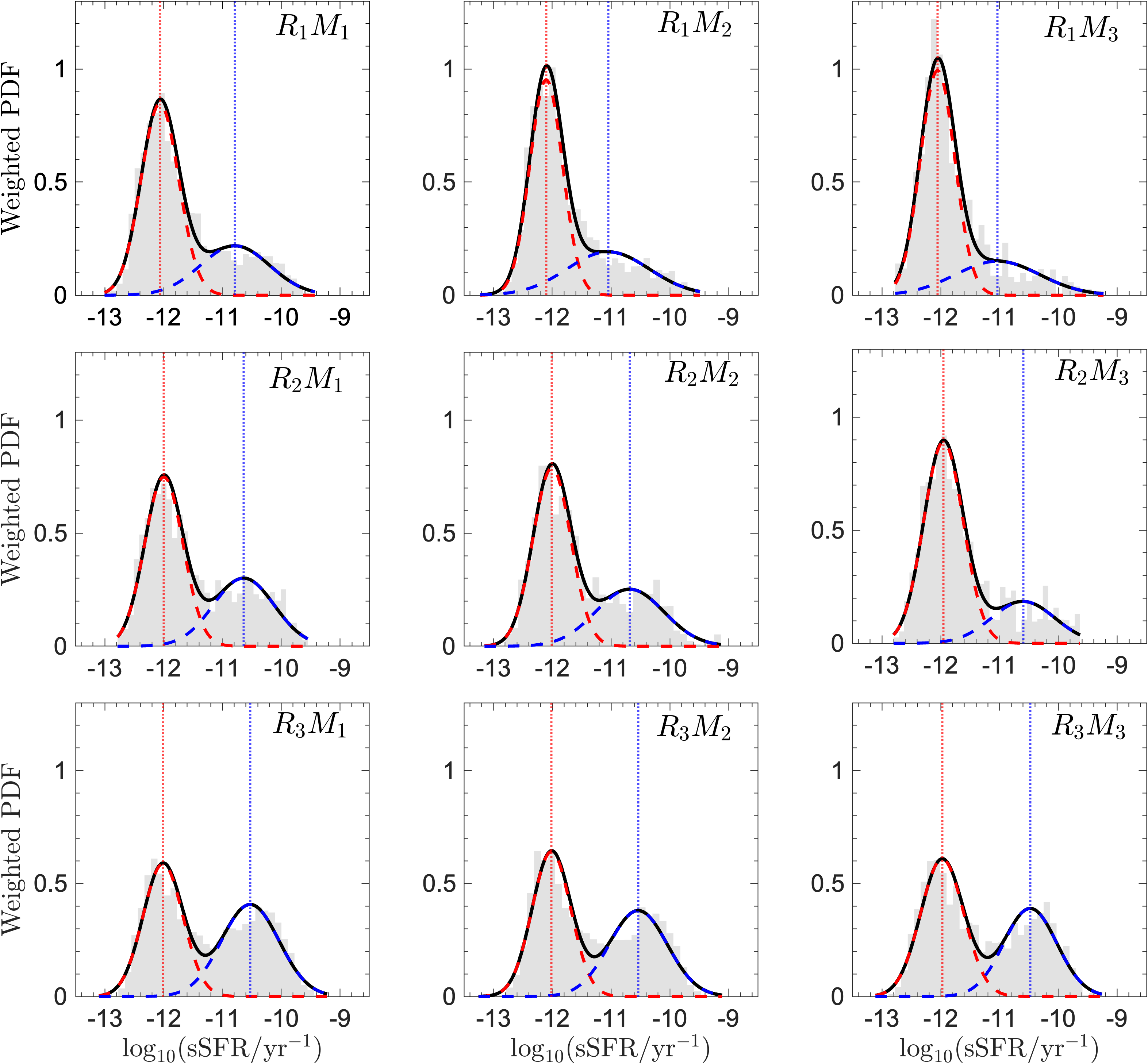} 
    \vspace{-0.5cm}
\caption{
Weighted $\log_{10}(\mathrm{sSFR}/\mathrm{yr}^{-1})$ distributions for the joint radial and host-halo-mass environments. Rows correspond to increasing clustercentric radius from the inner region, $R_1$, to the outskirts, $R_3$, while columns correspond to increasing host halo mass from $M_1$ to $M_3$. The gray histograms show the weighted observed distributions. The red and blue dashed curves represent the quenched and star-forming Gaussian components, respectively, and the solid black curve shows their sum. Vertical red and blue dotted lines mark the fitted component means, $\mu_{\rm Q}$ and $\mu_{\rm SF}$.
}
\label{fig:RM}
\end{figure*}
\begin{figure*}
\centering
\includegraphics[width=1\linewidth]{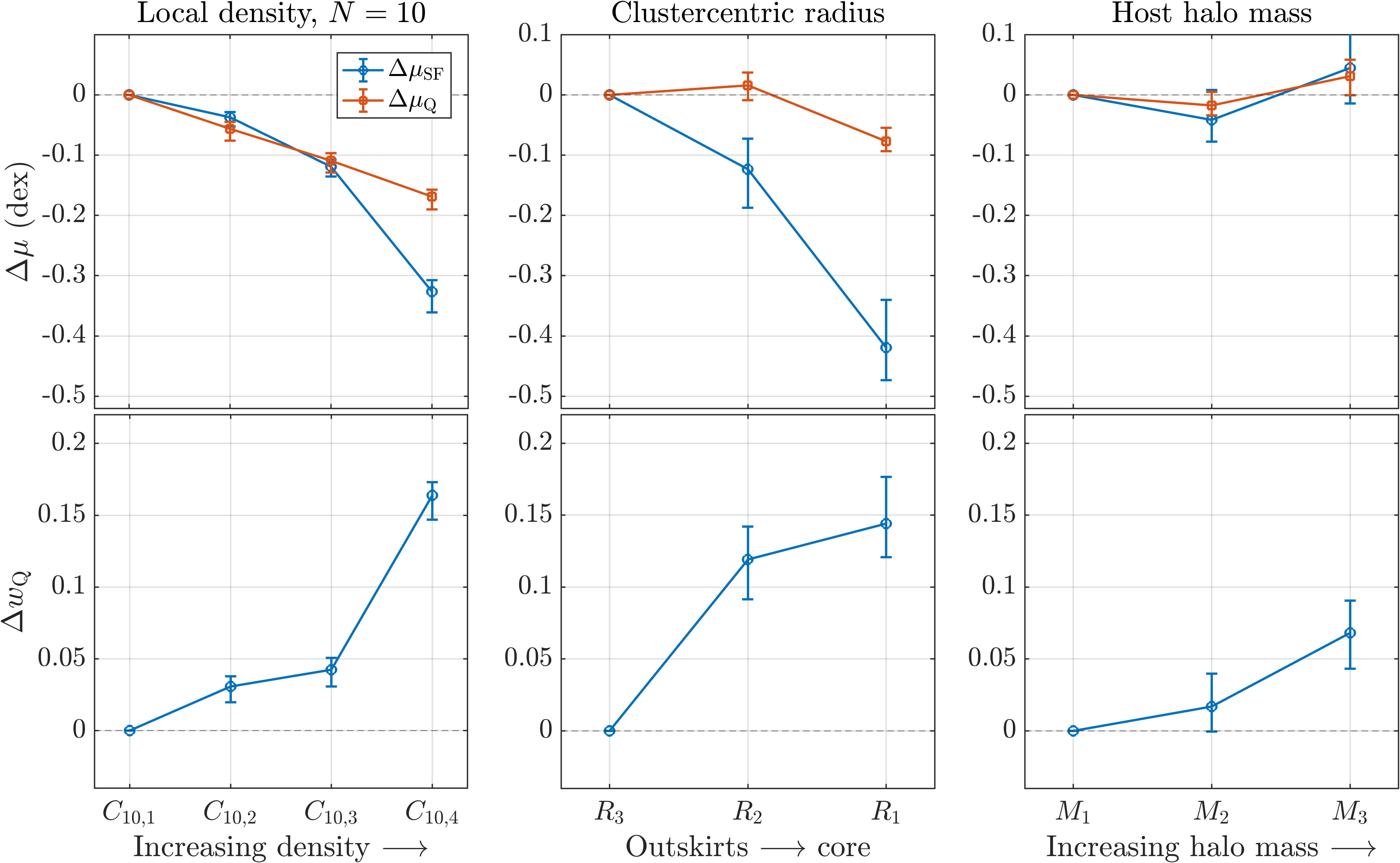} 
    \vspace{-0.5cm}
\caption{
Environmental changes in the Gaussian-mixture parameters relative to the least environmentally affected reference class. The left column shows the fiducial local-density classification, using $N=10$, relative to $C_{10,1}$. The middle column shows changes with clustercentric radius relative to the cluster-outskirts sample, $R_3$, and the right column shows changes with host halo mass relative to $M_1$. The upper panels show $\Delta\mu_{\rm SF}$ and $\Delta\mu_{\rm Q}$, while the lower panels show $\Delta w_{\rm Q}$. The reference points are fixed to zero by construction. The star-forming mean shifts toward lower sSFR with increasing local density and decreasing clustercentric radius, whereas the dependence on host halo mass is comparatively weak.
}
\label{fig:DeltaEnv}
\end{figure*}
\begin{figure*}
\centering
\includegraphics[width=1\linewidth]{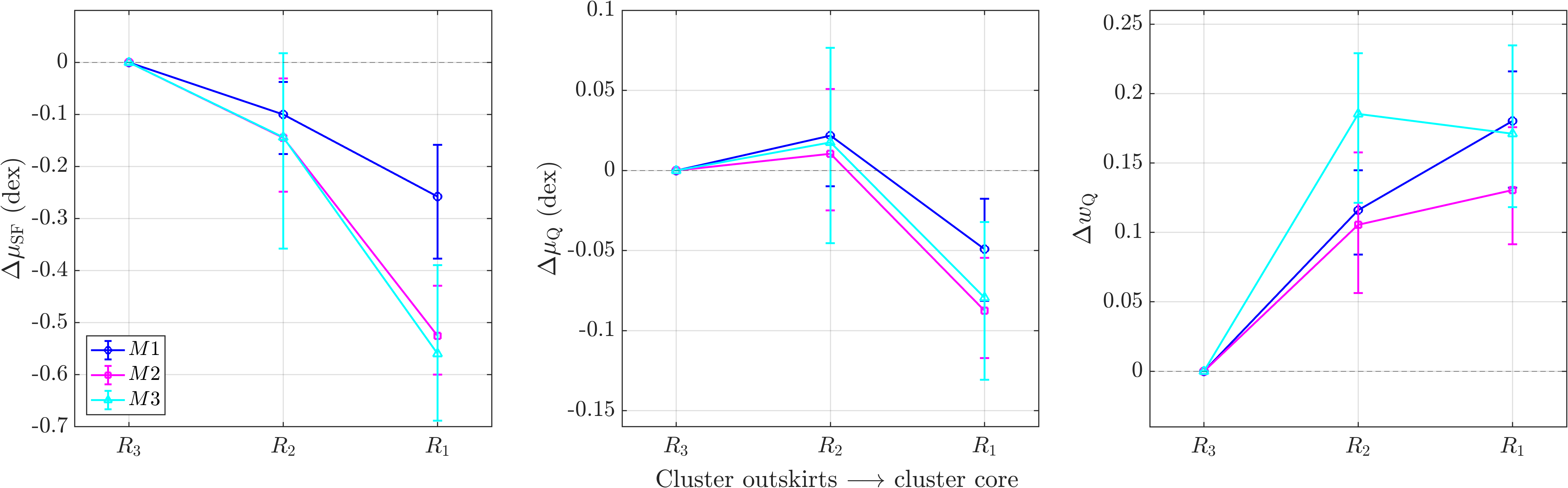} 
    \vspace{-0.5cm}
\caption{
Radial changes in the Gaussian-mixture parameters at fixed host halo mass. For each halo-mass class $M_j$, the quantities are measured relative to the corresponding cluster-outskirts sample, $R_3M_j$, such that $\Delta X(R_iM_j)=X(R_iM_j)-X(R_3M_j)$. The profiles are ordered from the  cluster outskirts and infall region, $R_3$, toward the cluster core, $R_1$. The left, middle, and right panels show $\Delta\mu_{\rm SF}$, $\Delta\mu_{\rm Q}$, and $\Delta w_{\rm Q}$, respectively. The reference points are fixed to zero by construction.
At fixed halo mass, the star-forming component shifts substantially toward lower sSFR toward the cluster center, while the quenched component mean changes comparatively little.
}
\label{fig:DeltaEnv_RM}
\end{figure*}

Figure~\ref{fig:DistLS} shows the weighted $\log_{10}(\mathrm{sSFR}/\mathrm{yr}^{-1})$ distributions for the four statistical density classes defined using the fiducial $\Sigma_{10}$ estimator; the corresponding distributions for the cluster-based
environments are shown in Figures~\ref{fig:DistSS} and~\ref{fig:RM} and discussed in Section~\ref{sec:DistSS}. A clear bimodal structure is present in all four classes, consisting of a lower-sSFR quenched component and a higher-sSFR star-forming component. However, their relative contributions change systematically with environment. From the lowest-density class, $C_{10,1}$, to the highest-density class, $C_{10,4}$, the quenched component becomes progressively more prominent. Its mixture weight increases from $w_{\rm Q}=0.46$ to $w_{\rm Q}=0.62$. This behavior is consistent with the increase in the independently measured weighted quenched fraction from $F_{\rm Q,w}=0.45$ in $C_{10,1}$ to $F_{\rm Q,w}=0.70$ in $C_{10,4}$, as reported in Table~\ref{tab:gmm_env}.

The left column of Figure~\ref{fig:DeltaEnv} quantifies the accompanying changes in the Gaussian-component means and quenched-component weight relative to the lowest density class, $C_{10,1}$. The peak of the quenched component changes comparatively modestly across the density sequence, shifting from $\mu_{\rm Q}\simeq-11.85$ in $C_{10,1}$ to $\mu_{\rm Q}\simeq-12.02$ in $C_{10,4}$, corresponding to $\Delta\mu_{\rm Q}\simeq-0.17$ dex. By contrast, the star-forming peak shifts more strongly toward lower sSFR, from $\mu_{\rm SF}\simeq-10.30$ to $\mu_{\rm SF}\simeq-10.62$, corresponding to $\Delta\mu_{\rm SF}\simeq-0.33$ dex. The lower-left panel further shows an increase of $\Delta w_{\rm Q}\simeq0.16$ between $C_{10,1}$ and $C_{10,4}$.

The same qualitative behavior is recovered for the $\Sigma_5$ and $\Sigma_{15}$ estimators. For each value of $N$, the highest-density class has the largest quenched-component weight and the lowest star-forming peak. The agreement among the three nearest-neighbor estimators indicates that the inferred environmental trends are not sensitive to the precise smoothing scale used to characterize the statistical density. Thus, the environmental dependence of the sSFR distribution is not limited to a change in the relative contributions of quenched and star-forming galaxies. Increasing local density is associated both with a larger quenched component and with a systematic shift of the star-forming component toward lower sSFR. This is consistent with reduced star formation activity among galaxies that remain within the star-forming population in denser environments.
\subsection{The sSFR Distribution in Cluster-Scale Environments}
\label{sec:DistSS}

We next examine how the full $\log_{10}(\mathrm{sSFR})$ distribution changes within cluster-scale environments. Unlike the statistical-density classes discussed above, these samples are restricted to the \texttt{GalWCat19} cluster members and are classified by projected clustercentric radius, host halo mass, or their joint combinations (Table~\ref{tab:env_classes}).

Figure~\ref{fig:DistSS} shows the weighted $\log_{10}(\mathrm{sSFR}/\mathrm{yr}^{-1})$ distributions and the corresponding two-component Gaussian-mixture fits for the radial samples $R_1$--$R_3$ and the halo-mass samples $M_1$--$M_3$. The corresponding fitted parameters and uncertainties are listed in
Table~\ref{tab:gmm_env}. The bimodal structure is visible in all cluster-scale environments, with a lower-sSFR quenched component and a higher-sSFR star-forming component. However, the relative contributions and fitted peak locations vary across the radial and halo-mass classes. The radial classes show a clear variation in the relative importance of the two components. The inner cluster region, $R_1$, has a quenched mixture weight of $w_{\rm Q}=0.66$. Moving outward to $R_2$ and $R_3$, the quenched mixture weight decreases to $w_{\rm Q}=0.64$ and $w_{\rm Q}=0.52$, respectively. This radial behavior indicates that the quenched component becomes progressively more prominent toward the cluster center.

The star-forming component also changes substantially with radius. In the innermost region, $R_1$, the star-forming peak is located at $\mu_{\rm SF}\simeq-10.94$, whereas in the cluster outskirts and infall region, $R_3$, it is located at $\mu_{\rm SF}\simeq-10.52$. The middle column of Figure~\ref{fig:DeltaEnv} quantifies this change relative to $R_3$, giving $\Delta\mu_{\rm SF}\simeq-0.42$ dex in $R_1$. By comparison, the quenched-component mean changes by only $\Delta\mu_{\rm Q}\simeq-0.08$ dex. The lower-middle panel also shows that the quenched-component weight increases by $\Delta w_{\rm Q}\simeq0.15$ from $R_3$ to $R_1$. This indicates that star-forming galaxies near the cluster center have lower sSFR than star-forming galaxies at larger projected radii. The radial trend therefore reflects both an increasing contribution from the quenched population and reduced star formation activity among galaxies that remain associated with the star-forming component.

The lower panels of Figure~\ref{fig:DistSS} show the corresponding distributions for the halo-mass classes. Compared with the radial classes, the variation with host halo mass is weaker. The quenched mixture weight increases from $w_{\rm Q}=0.56$ in $M_1$ to $w_{\rm Q}=0.63$ in $M_3$, indicating a modest increase in the relative contribution of the quenched component toward higher host halo mass. However, the peak positions of both Gaussian components vary only mildly and non-monotonically across $M_1$--$M_3$. This behavior is also visible in the right column of Figure~\ref{fig:DeltaEnv}, which shows that the changes in $\mu_{\rm SF}$ and $\mu_{\rm Q}$ with halo mass are small compared with those measured across the radial sequence. These results suggest that, within the cluster-associated sample, projected clustercentric radius is associated with a stronger variation in the sSFR distribution than host halo mass.

Because each halo-mass class contains galaxies spanning the full radial range, the $M_1$--$M_3$ distributions do not isolate the effects of halo mass and clustercentric radius. We therefore further examine the nine joint radial--halo-mass classes, $R_iM_j$, shown in Figure~\ref{fig:RM}. Each row corresponds to a fixed radial class, while each column corresponds to a fixed halo-mass class. The bimodal structure remains visible in all nine environments, but the position and relative prominence of the star-forming component change strongly with radius. At fixed halo mass, the star-forming peak moves systematically toward lower sSFR from the  cluster outskirts and infall region, $R_3$, to the inner region, $R_1$. For example, $\mu_{\rm SF}$ changes from $-10.53$ in $R_3M_1$ to $-10.79$ in $R_1M_1$, from $-10.54$ in $R_3M_2$ to $-11.06$ in $R_1M_2$, and from $-10.48$ in $R_3M_3$ to $-11.04$ in $R_1M_3$.

Figure~\ref{fig:DeltaEnv_RM} quantifies these radial changes relative to the corresponding outskirts sample, $R_3M_j$. The shift in the star-forming peak from $R_3$ to $R_1$ is $\Delta\mu_{\rm SF}\simeq-0.26$, $-0.53$, and $-0.56$ dex for $M_1$, $M_2$, and $M_3$, respectively. By contrast, the corresponding changes in the quenched-component mean are much smaller, approximately $-0.05$, $-0.09$, and $-0.08$ dex. The quenched-component weight also generally increases toward smaller radii, with net changes from $R_3$ to $R_1$ of $\Delta w_{\rm Q}\simeq0.18$, $0.13$, and $0.17$ for $M_1$, $M_2$, and $M_3$, respectively. For the highest-mass class, the quenched weight reaches its largest value in $R_2$ rather than $R_1$, so this particular trend is not strictly monotonic.

At fixed clustercentric radius, the variations across the halo-mass classes are generally weaker and less systematic than the radial changes. This is particularly clear in the $R_2$ and $R_3$ samples, although the inner $R_1$ region shows a larger difference between the $M_1$ class and the two higher-mass classes. The joint classification therefore confirms that the strong inward shift of the star-forming component is not attributable solely to differences in the halo-mass distributions of the radial samples. Instead, galaxies closer to the cluster center exhibit lower star formation activity even within fixed host halo-mass classes.

Overall, the cluster-scale analysis shows that projected position within the cluster is associated with a substantially stronger variation in the sSFR distribution than host halo mass alone. The inner cluster regions contain a larger quenched component and exhibit reduced star formation activity among galaxies that remain associated with the star-forming population. The persistence of these trends within fixed halo-mass classes indicates that clustercentric radius, and therefore a galaxy's location within the cluster environment, is more strongly associated with its current star formation activity than host halo mass.
\begin{figure*}\centering
\includegraphics[width=0.8\linewidth]{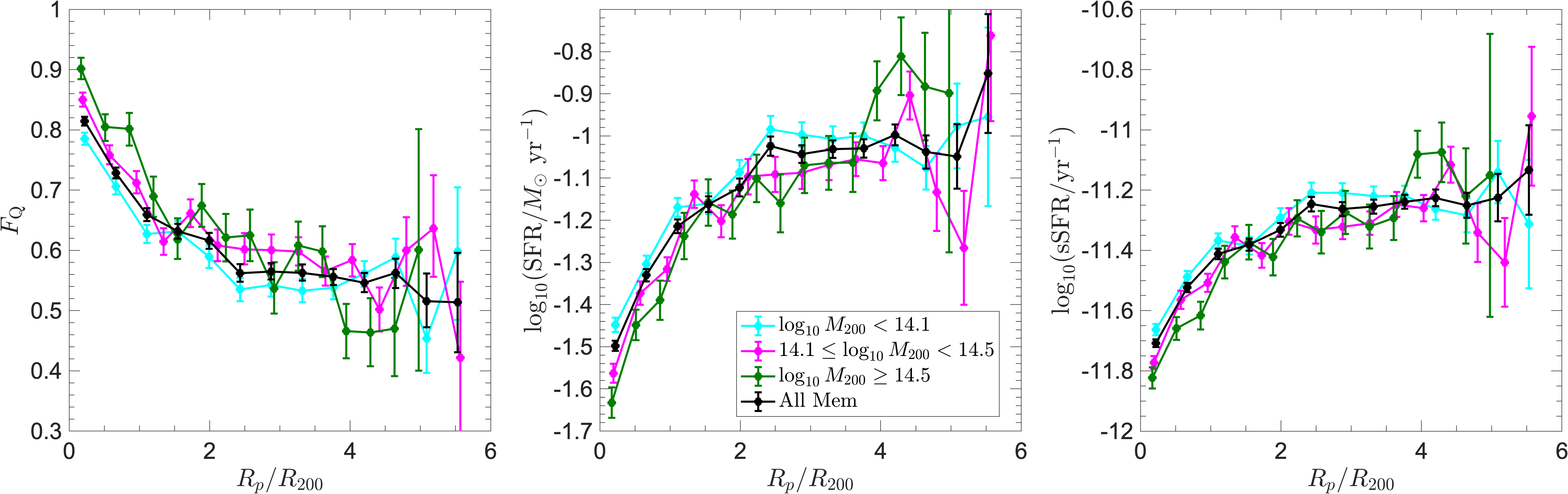} 
    \vspace{-0cm}
\caption{
Radial dependence of star formation activity for galaxies associated with \texttt{GalWCat19} clusters. From left to right, the panels show the weighted quenched fraction, $F_{\rm Q}$, the weighted mean $\log_{10}(\mathrm{SFR}/M_\odot~\mathrm{yr}^{-1})$, and the weighted mean $\log_{10}(\mathrm{sSFR}/\mathrm{yr}^{-1})$ as functions of projected clustercentric radius, $R_{\rm p}/R_{200}$. Colored curves show the three host halo-mass bins, while the black curve shows the full cluster-associated sample. Error bars show the $1\sigma$ uncertainties estimated from bootstrap resampling.
}
\label{fig:radial}
\end{figure*}
\begin{figure*}\centering
\includegraphics[width=0.7\linewidth]{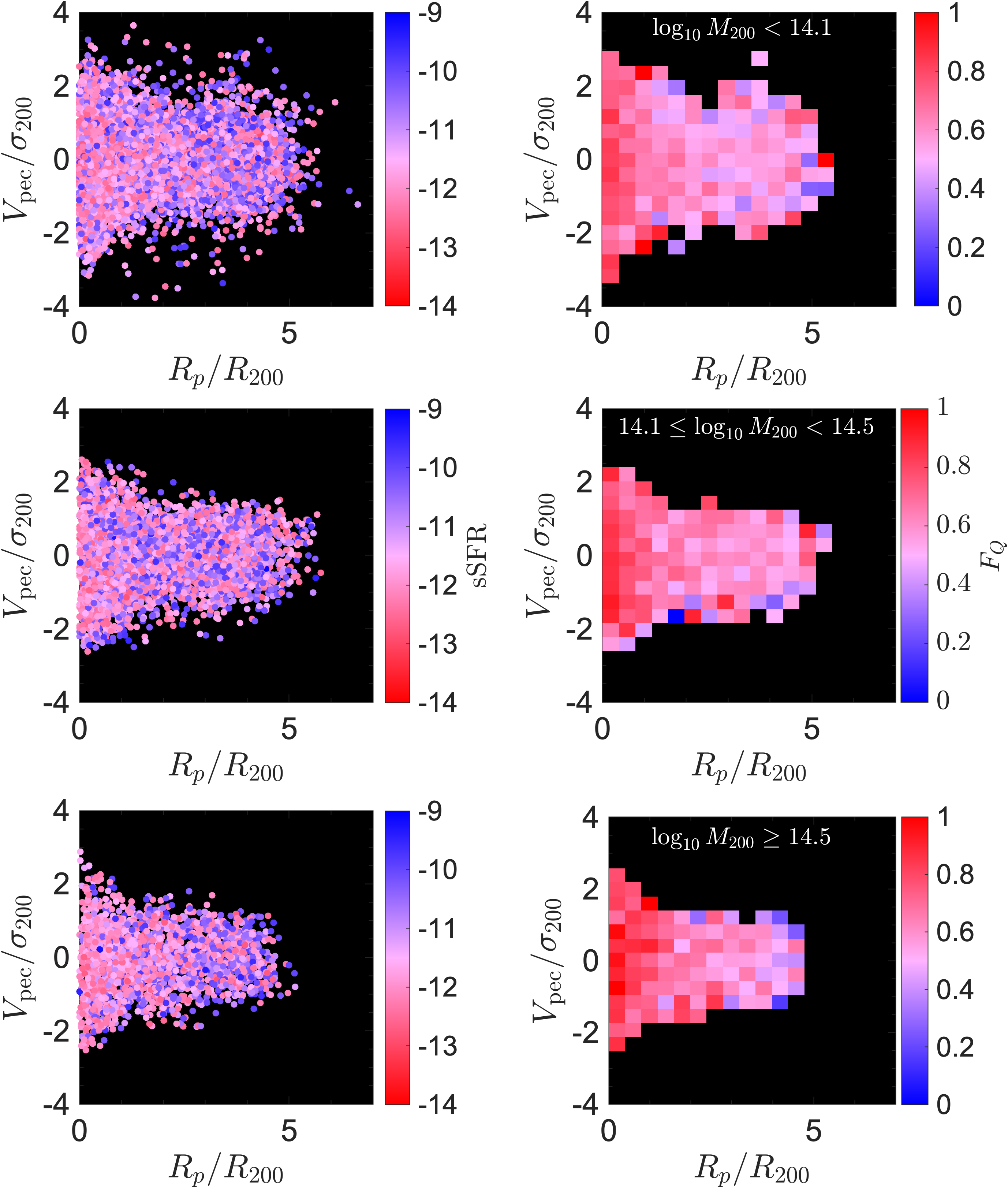} 
    \vspace{-0.25cm}
\caption{
Projected phase-space distributions of galaxies associated with \texttt{GalWCat19} clusters, separated by host halo mass. The horizontal axis shows the projected clustercentric radius, $R_{\rm p}/R_{200}$, and the vertical axis shows the normalized line-of-sight peculiar velocity, $V_{\rm pec}/\sigma_{200}$. From top to bottom, the rows correspond to the halo-mass bins $M_1$, $M_2$, and $M_3$. In the left column, individual galaxies are colored by $\log_{10}(\mathrm{sSFR}/\mathrm{yr}^{-1})$. In the right column, colors show the weighted quenched fraction, $F_{\rm Q}$, in each projected phase-space bin. 
} 
\label{fig:phase_space}
\end{figure*}
\subsection{Star Formation Activity as a Function of Clustercentric Radius}
\label{sec:cluster_radial_trends}

The Gaussian-mixture analysis above shows that the sSFR distribution varies most strongly with projected clustercentric radius. We therefore examine this radial dependence in greater detail by measuring the weighted quenched fraction, the weighted mean $\log_{10}(\mathrm{SFR}/M_\odot~\mathrm{yr}^{-1})$, and the weighted mean $\log_{10}(\mathrm{sSFR}/\mathrm{yr}^{-1})$ as functions of $R_{\rm p}/R_{200}$. To test whether these radial trends vary with host halo mass, we calculate the same quantities separately for the three halo-mass classes defined in Table~\ref{tab:env_classes}, as well as for the full cluster-associated sample.

Figure~\ref{fig:radial} shows the resulting radial profiles. The left panel shows that the weighted quenched fraction generally decreases with increasing projected clustercentric radius. The highest values of $F_{\rm Q}$ are found in the inner cluster regions, where $F_{\rm Q}\gtrsim0.8$ for the full cluster-associated sample and for the individual halo-mass classes $M_1$, $M_2$, and $M_3$. At larger projected radii, $F_{\rm Q}$ declines to values of approximately $0.5$--$0.6$ in the cluster outskirts and infall region. This radial behavior shows that galaxies close to the cluster center are more frequently quenched than galaxies at larger projected radii.

The inner cluster region also shows a tendency for the quenched fraction to increase with host halo mass, from the lower-mass class, $M_1$, to the highest-mass class, $M_3$. This suggests that the association between cluster environment and suppressed star formation is strongest in the central regions of the most massive halos.

The middle panel shows the corresponding radial behavior of the weighted mean $\log_{10}(\mathrm{SFR})$. In all three halo-mass classes, the mean star formation activity is lowest near the cluster center and increases toward larger projected radii. Within the inner cluster region, galaxies in the most massive clusters generally have lower mean $\log_{10}(\mathrm{SFR})$ than galaxies in lower-mass systems. The halo-mass behavior seen in $F_{\rm Q}$ is therefore also reflected in the absolute star formation activity of the galaxy population.

The right panel shows a similar radial trend in the weighted mean $\log_{10}(\mathrm{sSFR})$. The mean sSFR increases outward in all halo-mass classes, indicating that the radial trend is not driven solely by differences in stellar mass. Instead, galaxies at larger projected radii are also forming stars more actively relative to their already assembled stellar mass. As in the SFR profile, the inner regions of the most massive clusters generally show lower mean sSFR than the corresponding regions of lower-mass clusters.

Overall, the radial profiles are consistent with the Gaussian-mixture results. The inner cluster regions contain the largest quenched fractions and the lowest weighted mean $\log_{10}(\mathrm{SFR})$ and $\log_{10}(\mathrm{sSFR})$, whereas galaxies in the outskirts and infall region show higher star formation activity and a lower incidence of quenching. The dependence on host halo mass is most apparent near the cluster center and becomes less distinct at larger projected radii.
These results show that projected clustercentric radius is the dominant cluster-scale variable associated with the current star formation activity of cluster members, while host halo mass provides a weaker secondary dependence.
\begin{table*}
\centering
\caption{Environmental quenching efficiency for the cluster radial and host halo-mass classes.}
\label{tab:Eff}
\begin{tabular}{lcccc}
\hline
\hline
Environment &
$\log_{10}(M_\star/h^{-2}M_\odot)$ &
$F_{\rm Q,ref}$ &
$F_{\rm Q,env}$ &
$\epsilon_{\rm env}$ \\
\hline

\multicolumn{5}{l}{\textit{Clustercentric-radius classes}} \\

$R_1:\ R_{\rm p}/R_{200}<0.5$
& $9.7\text{--}10.0$  & 0.298 & 0.755 & $0.651^{+0.017}_{-0.018}$ \\
& $10.0\text{--}10.3$ & 0.449 & 0.801 & $0.639^{+0.023}_{-0.021}$ \\
& $10.3\text{--}10.6$ & 0.614 & 0.851 & $0.616^{+0.033}_{-0.030}$ \\
& $10.6\text{--}11.0$ & 0.788 & 0.947 & $0.750^{+0.047}_{-0.047}$ \\

\hline
$R_2:\ 0.5\leq R_{\rm p}/R_{200}<1.0$
& $9.7\text{--}10.0$  & 0.298 & 0.642 & $0.490^{+0.024}_{-0.023}$ \\
& $10.0\text{--}10.3$ & 0.449 & 0.708 & $0.470^{+0.026}_{-0.029}$ \\
& $10.3\text{--}10.6$ & 0.614 & 0.809 & $0.507^{+0.041}_{-0.038}$ \\
& $10.6\text{--}11.0$ & 0.788 & 0.870 & $0.389^{+0.095}_{-0.084}$ \\

\hline
$R_3:\ 1.0\leq R_{\rm p}/R_{200}\leq6.0$
& $9.7\text{--}10.0$  & 0.298 & 0.445 & $0.210^{+0.013}_{-0.014}$ \\
& $10.0\text{--}10.3$ & 0.449 & 0.572 & $0.224^{+0.018}_{-0.018}$ \\
& $10.3\text{--}10.6$ & 0.614 & 0.714 & $0.261^{+0.025}_{-0.029}$ \\
& $10.6\text{--}11.0$ & 0.788 & 0.852 & $0.301^{+0.052}_{-0.060}$ \\

\hline
\multicolumn{5}{l}{\textit{Host halo-mass classes}} \\

$M_1:\ \log_{10}(M_{200}/h^{-1}M_\odot)<14.1$
& $9.7\text{--}10.0$  & 0.298 & 0.518 & $0.313^{+0.015}_{-0.014}$ \\
& $10.0\text{--}10.3$ & 0.449 & 0.623 & $0.317^{+0.018}_{-0.020}$ \\
& $10.3\text{--}10.6$ & 0.614 & 0.744 & $0.338^{+0.027}_{-0.028}$ \\
& $10.6\text{--}11.0$ & 0.788 & 0.875 & $0.408^{+0.052}_{-0.056}$ \\

\hline
$M_2:\ 14.1\leq\log_{10}(M_{200}/h^{-1}M_\odot)<14.5$
& $9.7\text{--}10.0$  & 0.298 & 0.572 & $0.390^{+0.018}_{-0.015}$ \\
& $10.0\text{--}10.3$ & 0.449 & 0.655 & $0.374^{+0.020}_{-0.021}$ \\
& $10.3\text{--}10.6$ & 0.614 & 0.774 & $0.414^{+0.034}_{-0.029}$ \\
& $10.6\text{--}11.0$ & 0.788 & 0.882 & $0.443^{+0.061}_{-0.062}$ \\

\hline
$M_3:\ \log_{10}(M_{200}/h^{-1}M_\odot)\geq14.5$
& $9.7\text{--}10.0$  & 0.298 & 0.595 & $0.423^{+0.024}_{-0.029}$ \\
& $10.0\text{--}10.3$ & 0.449 & 0.672 & $0.405^{+0.030}_{-0.030}$ \\
& $10.3\text{--}10.6$ & 0.614 & 0.757 & $0.372^{+0.046}_{-0.050}$ \\
& $10.6\text{--}11.0$ & 0.788 & 0.878 & $0.423^{+0.093}_{-0.105}$ \\
\hline
\end{tabular}

\tablecomments{
The lowest-density statistical class, $C_{10,1}$, which contains no
\texttt{GalWCat19} members, is adopted as the low-density non-cluster reference
population. Here, $F_{\rm Q,ref}$ and $F_{\rm Q,env}$ are the
$1/V_{\max}$-weighted quenched fractions in the reference and target
environments, respectively. The environmental quenching efficiency is
defined as
$\epsilon_{\rm env}=(F_{\rm Q,env}-F_{\rm Q,ref})/
(1-F_{\rm Q,ref})$.
Quoted uncertainties correspond to the 16th and 84th percentiles of the
bootstrap distribution.
}
\end{table*}

\begin{figure*}\centering
\includegraphics[width=0.8\linewidth]{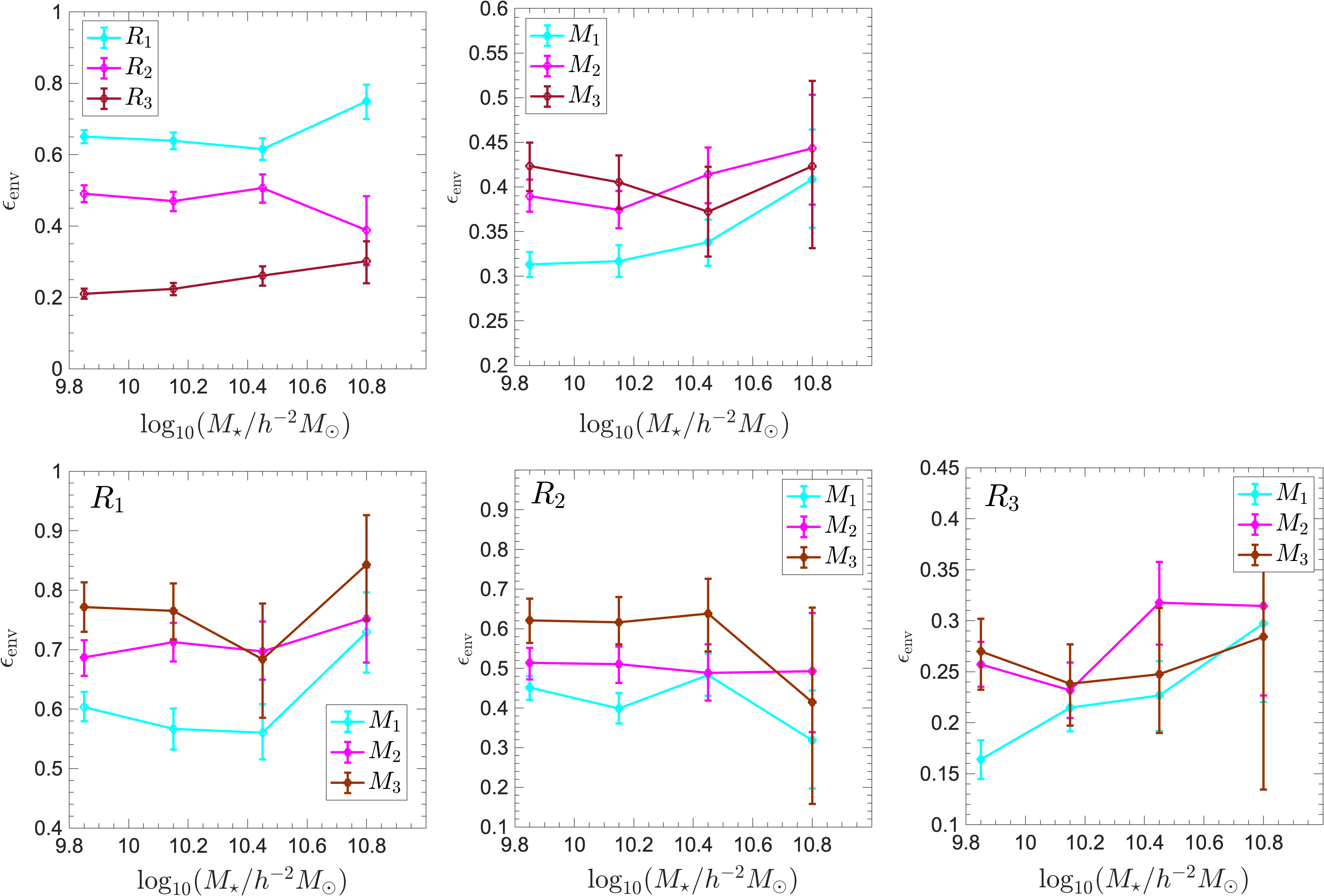} 
    \vspace{-0cm}
\caption{Environmental quenching efficiency, $\epsilon_{\rm env}$, as a function of stellar mass for the cluster radial and host halo-mass subsamples, using the lowest-density statistical class, $C_{10,1}$, as the reference population. The top-left panel shows the radial classes $R_1$--$R_3$, while the top-right panel shows the host halo-mass classes $M_1$--$M_3$. The bottom row isolates the dependence on host halo mass at fixed projected clustercentric radius, with $R_1$, $R_2$, and $R_3$ shown from left to right. Each bottom panel compares $M_1$, $M_2$, and $M_3$. Error bars correspond to the 16th and 84th percentiles of the bootstrap distributions. }
\label{fig:Eff}
\end{figure*}
\subsection{Star Formation Activity in Projected Phase Space}
\label{sec:phase_space}

We next examine the distribution of star formation activity in projected phase space for galaxies associated with \texttt{GalWCat19} clusters. The projected phase space is defined using the projected clustercentric radius, $R_{\rm p}/R_{200}$, and the normalized line-of-sight peculiar velocity, $V_{\rm pec}/\sigma_{200}$. By combining projected position and line-of-sight velocity, this representation provides a more detailed characterization of the cluster environment than projected radius alone.

Figure~\ref{fig:phase_space} shows the projected phase-space distributions for the three host halo-mass bins. From top to bottom, the rows correspond to the halo-mass bins $M_1$, $M_2$, and $M_3$, respectively. In the left column, individual galaxies are colored by their $\log_{10}(\mathrm{sSFR}/\mathrm{yr}^{-1})$, while the right column shows the weighted quenched fraction, $F_{\rm Q}$, measured in bins of projected phase space. The individual-galaxy distributions span a broad range of normalized peculiar velocities near the cluster center. The maximum observed absolute normalized peculiar velocity, $|V_{\rm pec}|/\sigma_{200}$, decreases with increasing projected clustercentric radius, producing the characteristic trumpet-shaped boundary in projected phase space. This radial narrowing of the velocity distribution is visible in all three halo-mass bins. Galaxies with the lowest sSFR are preferentially concentrated toward small $R_{\rm p}/R_{200}$, while galaxies with higher sSFR become more common at larger projected radii. This behavior is consistent with the radial trends shown in Figure~\ref{fig:radial}, where the mean sSFR increases outward from the cluster center.

The phase-space maps of $F_{\rm Q}$ show that the highest quenched fractions occur predominantly in the inner cluster regions. In all three halo-mass bins, the central phase-space region is dominated by quenched galaxies, whereas lower quenched fractions are generally found at larger projected radii. This pattern is consistent with a population in the inner cluster regions that has experienced stronger or more prolonged environmental processing, while the outskirts and infall regions contain a larger fraction of recently accreted galaxies that have not yet fully quenched. The largest quenched fractions are concentrated toward the central region of projected phase space, where galaxies are expected to have experienced the cluster environment for longer periods, consistent with previous studies \citep{Barsanti18,Abdullah26Coma}.

A dependence on host halo mass is also apparent. To quantify this dependence directly in projected phase space, we compare the $M_1$, $M_2$, and $M_3$ samples cell by cell using an identical phase-space grid and retain only cells containing at least 10 galaxies in each halo-mass sample. Across these common cells, $67\%$ have a higher quenched fraction in $M_3$ than in $M_1$, with a median difference of $\Delta F_{\rm Q}=F_{\rm Q}(M_3)-F_{\rm Q}(M_1)=0.066$. The halo-mass dependence varies with projected radius. Within $R_1$, $74\%$ of the common cells have $F_{\rm Q}(M_3)>F_{\rm Q}(M_1)$, increasing to $83\%$ in $R_2$, compared with $61\%$ in $R_3$. The corresponding median differences in quenched fraction are $0.086$, $0.132$, and $0.034$, respectively. These results quantitatively support a secondary dependence on host halo mass, which is most clearly expressed through the quenched fraction within $R_{200}$.
\subsection{Environmental Quenching Efficiency}
\label{sec:quenching_efficiency}

To distinguish environmental quenching from quenching linked to internal galaxy properties, particularly stellar mass, we calculate the environmental quenching efficiency in bins of stellar mass. 

We adopt $C_{10,1}$, which contains no galaxies identified as
\texttt{GalWCat19} members, as the low-density non-cluster reference
population. We define the environmental quenching efficiency as \citep[e.g.,][]{Peng10,vanDerBurg18}
\begin{equation}
\epsilon_{\rm env}(M_\star,E)
=
\frac{
F_{\rm Q}(M_\star,E)
-
F_{\rm Q}(M_\star,C_{10,1})
}{
1-F_{\rm Q}(M_\star,C_{10,1})
},
\end{equation}
where $F_{\rm Q}(M_\star,E)$ is the weighted quenched fraction in a given cluster environment, $E$, and $F_{\rm Q}(M_\star,C_{10,1})$ is the corresponding weighted quenched fraction in the reference environment at the same stellar mass. All quenched fractions are calculated using the adopted $1/V_{\max}$ galaxy weights.

This quantity represents the excess quenched fraction in the target environment, normalized by the fraction of galaxies that remain star-forming in the reference environment at the same stellar mass. Thus, $\epsilon_{\rm env}=0$ indicates no excess quenching relative to $C_{10,1}$, whereas $\epsilon_{\rm env}>0$ indicates an enhanced probability of quenching in the cluster environment. By calculating $\epsilon_{\rm env}$ separately within narrow stellar-mass bins, we compare galaxies with similar stellar masses across different environments. This reduces the possibility that an environment appears more strongly quenched simply because it contains a larger fraction of massive galaxies, which are intrinsically more likely to be quiescent. 

The resulting excess quenched fraction can therefore be interpreted as an estimate of the environmental contribution relative to the adopted low-density reference population at fixed stellar mass. The weighted quenched fractions and corresponding environmental quenching efficiencies for the radial and global halo-mass subsamples are listed in Table~\ref{tab:Eff}. Figure~\ref{fig:Eff} shows $\epsilon_{\rm env}$ as a function of stellar mass for the cluster radial and host halo-mass subsamples. The top-left panel shows a strong dependence on projected clustercentric radius. In every stellar-mass bin, the inner cluster region, $R_1$, has the highest quenching efficiency, the outer virial region, $R_2$, has intermediate values, and the outskirts and infall region, $R_3$, has the lowest values.  The ordering $\epsilon_{\rm env}(R_1)>\epsilon_{\rm env}(R_2)>\epsilon_{\rm env}(R_3)$ is maintained across the full stellar-mass range considered here. This indicates that the probability of excess quenching increases strongly toward the cluster center. Moreover, the differences among the radial classes are substantially larger than the variation with stellar mass within an individual radial class, showing that projected clustercentric radius is the dominant variable associated with the environmental quenching efficiency.

The top-right panel shows the corresponding efficiencies for the three host halo-mass bins after combining galaxies over the full radial range. The differences among $M_1$, $M_2$, and $M_3$ are substantially weaker than the radial dependence. However, each halo-mass bin contains galaxies spanning the cluster core, virial region, outskirts, and infall region. Because the quenching efficiency varies strongly with clustercentric radius, these measurements represent averages over different radial environments and do not isolate the effect of host halo mass alone. A more direct assessment therefore requires comparing the halo-mass bins at fixed stellar mass and fixed projected clustercentric radius.

The bottom row performs this controlled comparison by showing $M_1$, $M_2$, and $M_3$ separately within each radial class. In the inner cluster region, $R_1$, the environmental quenching efficiency generally increases with host halo mass, particularly over the three lowest stellar-mass bins. This indicates that, at fixed galaxy stellar mass and clustercentric radius, quenching is more efficient in the cores of more massive clusters. A similar halo-mass dependence is present in the outer virial region, $R_2$, where $M_3$ generally has the highest quenching efficiency and $M_1$ the lowest, although the differences are smaller than those measured in the cluster core. In the outskirts and infall region, $R_3$, the efficiencies of the three halo-mass classes are more similar and do not follow a consistent ordering. Their bootstrap intervals also overlap substantially, indicating that the halo-mass dependence becomes weak or statistically unclear at large projected radii. The highest-stellar-mass measurements should be interpreted cautiously because their bootstrap uncertainties are larger.

Overall, the environmental quenching efficiency is positive across all radial and halo-mass classes, demonstrating that cluster members experience excess quenching relative to the lowest-density reference population even after controlling for stellar mass. The strongest dependence is on projected clustercentric radius: quenching is most efficient in the cluster core and decreases systematically toward the outskirts and infall regions. Host halo mass introduces a weaker secondary dependence, with more efficient quenching generally found in higher-mass systems, particularly in the inner cluster regions. These results indicate that clustercentric radius is the dominant cluster-scale variable associated with environmental quenching, while host halo mass modulates the strength of this effect most clearly near the cluster center.
\section{Discussion and Conclusions}
\label{sec:discussion}
We investigated how galaxy star formation activity depends on local projected density, projected clustercentric radius, and host halo mass at low redshift. A central result of this work is that environmental suppression of star formation is not expressed solely through an increasing quenched fraction. The characteristic sSFR of the star-forming component decreases by approximately 0.29--0.35 dex from the lowest- to highest-density environments and by approximately 0.42 dex from the cluster outskirts to the inner cluster region. Thus, galaxies that remain star forming already exhibit systematically reduced star formation activity in denser environments and toward cluster centers.

This extends the picture emphasized by \citet{Wetzel12,Wetzel13}, in which environmental trends are primarily reflected in changes in the quenched fraction, with comparatively little modification of the star-forming population before quenching. Our measurements show that the influence of environment is also detectable within the star-forming population itself. Our results therefore reveal two complementary environmental trends: the star-forming population shifts toward lower sSFR, while the probability of galaxies being quenched increases. The environmental-quenching-efficiency measurements provide complementary evidence for the latter effect by demonstrating excess quenching in cluster environments at fixed stellar mass.

\subsection{Environmental Dependence of the sSFR Distribution}

The environmental dependence is evident in several complementary measures of star formation activity. The weighted mean SFR and sSFR decrease continuously with increasing $\Sigma_N$, while the quenched fraction increases. These trends are consistently recovered for $N=5$, 10, and 15, demonstrating that the measured dependence is robust to the adopted nearest-neighbor scale and is consistent with previous studies linking galaxy star formation to local density \citep{Lewis02,Balogh04,DePropris04,Peng10}.

The full sSFR distributions provide additional information beyond the change in quenched fraction. The Gaussian-mixture decomposition shows that changes in environment affect both the relative contributions of the quenched and star-forming populations and the location of the star-forming component itself. In particular, the quenched peak changes comparatively little, whereas the star-forming peak shifts systematically toward lower sSFR with increasing local density and decreasing clustercentric radius. This behavior cannot be explained simply by changing the relative numbers of star-forming and quenched galaxies; it indicates that star formation is already suppressed among galaxies that remain part of the star-forming population.

Such suppression may reflect environmental processes acting before complete quenching. Gradual mechanisms such as strangulation or starvation can reduce the available gas supply over extended timescales, while more rapid processes such as ram-pressure stripping may become increasingly effective in the densest cluster environments \citep{Larson80,Balogh00,Kauffmann04,Gunn72,Abadi99,Boselli19}. The present analysis does not uniquely distinguish among these mechanisms, which may operate together and over different timescales.

\subsection{Radial Dependence of Star Formation within Galaxy Clusters}

Within the \texttt{GalWCat19} cluster sample, the inner regions have the highest quenched fractions and the lowest mean SFRs and sSFRs, while star-forming galaxies become more common toward the outskirts. Projected clustercentric radius traces several environmental conditions simultaneously. Galaxies approaching the cluster center encounter a denser intracluster medium, stronger tidal fields, higher galaxy densities, and potentially longer cumulative exposure to the cluster environment. Ram-pressure stripping, strangulation, tidal interactions, and repeated high-speed encounters may therefore act together to produce the observed radial gradient. 

The projected phase-space distributions support this interpretation. Galaxies with low sSFR are preferentially concentrated toward small $R_{\rm p}/R_{200}$, whereas galaxies with higher sSFR become more common at larger projected radii. This statistical pattern is consistent with stronger or more prolonged environmental processing in the central regions and a larger contribution from recently accreted or infalling galaxies in the outskirts \citep{Barsanti18,Abdullah2026Coma}. However, projected phase-space position cannot uniquely determine the accretion history of individual galaxies.

Environmental quenching also remains measurable in the outskirts and infall region. This may indicate that environmental processing begins beyond the virial radius or that some galaxies are pre-processed in groups, filaments, or infalling structures before entering the main cluster \citep{Fujita04,McGee09,Haines15,Darvish17}. Complementary evidence at $z\sim1$ was presented by \citet{Werner22}, who found that many massive quiescent galaxies in cluster infall regions were likely self-quenched or pre-processed before cluster entry, whereas quenching after crossing the virial radius remained important for lower-mass galaxies. The broad $R_3$ region may contain a mixture of first-infalling, backsplash, pre-processed, and projected galaxies, which may contribute differently to the measured quenching signal.

\subsection{Environmental Quenching and Host Halo Mass}

The environmental-quenching-efficiency analysis complements the sSFR distribution results by quantifying excess quenching at fixed stellar mass. Environmental quenching efficiencies are positive in every stellar-mass and cluster-environment bin relative to the low-density non-cluster reference population. Cluster members therefore have a higher probability of being quenched than galaxies of the same stellar mass in the reference environment.
This demonstrates that the observed cluster trends cannot be explained by internal stellar-mass quenching alone and require an additional environmental contribution. At fixed stellar mass, excess quenching is strongest in the cluster core, intermediate in the outer virial region, and weakest in the outskirts and infall region. This radial ordering is maintained over the full stellar-mass range considered here, providing direct evidence that the environmental contribution increases toward the cluster center.

The dependence on host halo mass is weaker. When galaxies are combined over the full radial range, only modest differences are found among the halo-mass classes because each class mixes galaxies from the core, virial region, and outskirts. A clearer halo-mass dependence emerges after controlling for both stellar mass and projected clustercentric radius. Quenching is generally more efficient in the inner regions of higher-mass clusters, while the halo-mass dependence becomes weak or statistically unclear in the outskirts.

This behavior is physically plausible because more massive clusters generally contain hotter and denser intracluster media and have larger characteristic galaxy velocities. Processes related to the intracluster medium, particularly ram-pressure stripping, may therefore be more effective in their central regions. At larger radii, variations in orbital history, infall time, local substructure, and pre-processing may become more important than host halo mass. Our results therefore suggest that halo mass modulates the strength of environmental quenching, particularly near the cluster center, rather than determining the primary cluster-scale trend.

\subsection{Implications for Galaxy-Evolution Models}

The measured SFR and sSFR distributions provide quantitative constraints for simulations and galaxy-formation models. Reproducing only the total quenched fraction is insufficient to describe the observed environmental dependence of galaxy star formation. Successful models should simultaneously reproduce the bimodal shape of the sSFR distribution, the environmental variation in the quenched and star-forming components, and the shift of the star-forming peak toward lower sSFR in denser and more central environments.

Models should also recover the strong radial gradients in the quenched fraction, mean SFR, and mean sSFR, together with the weaker halo-mass dependence that becomes most apparent near the cluster center. Comparisons with simulations should apply observationally consistent selections, including projected clustercentric radii, line-of-sight velocities, stellar-mass limits, and cluster membership criteria. Such comparisons can test whether simulated quenching timescales and environmental mechanisms produce the observed spatial and phase-space distributions of star-forming and quenched cluster members.

Because both samples are based on SDSS spectroscopy, fiber collisions may preferentially miss galaxies in dense regions, leading to underestimated local densities and incomplete cluster-member sampling near cluster centers. Nevertheless, the environmental trends are recovered consistently across the nearest-neighbor estimators and toward smaller clustercentric radii. Spectroscopic incompleteness is therefore unlikely to change our qualitative conclusions, although it may affect the measured amplitudes of these trends.

In summary, the principal result of this work is that environmental suppression of star formation is already detectable within the star-forming galaxy population, rather than appearing only through an increase in the quenched fraction. Environmental-quenching-efficiency measurements provide complementary evidence that the same environments also increase the probability of galaxies becoming quenched, even at fixed stellar mass. Local density and projected clustercentric radius show the strongest associations with both effects, while host halo mass provides a weaker secondary dependence, most apparent near cluster centers. Together, these results show that environmental influence is evident both in the star formation activity of galaxies that remain star forming and in the probability that galaxies are quenched.

\section*{Acknowledgements}
GW gratefully acknowledges support from the National Science Foundation through grant AST-2347348.





\begin{acknowledgments}
\end{acknowledgments}

\bibliography{Ref}{}
\bibliographystyle{aasjournal}

\end{document}